\documentstyle[useAMS,times,psfig,referee]{mn2e}

\newif\ifAMStwofonts

\newcommand{\beq}{\begin{equation}}
\newcommand{\eeq}{\end{equation}}
\newcommand{\roma}[1] {{\rm #1}}
\newcommand{\Msun}{\hbox{${\rm M}_{\odot}$}}
\newcommand{\Lsun}{\hbox{${\rm L}_{\odot}$}}

\newcommand{\Mc}{\mbox{$M_\roma{c}$}}
\newcommand{\Teff}{\mbox{$T_\roma{eff}$}}
\newcommand{\RZAMS}{\mbox{$R_\roma{\rm zams}$}}
\newcommand{\Menv}{\mbox{$M_{\rm env}$}}
\input{psfig.sty}

\newif\ifAMStwofonts
\def\Mesz{M\'esz\'aros~}

\title[Core collapse SNe and GRBs]
      {Formation rates of core collapse  SNe and GRBs}
      \author[R.~G. Izzard, E. Ramirez-Ruiz \& C.~A. Tout ] {Robert
      G. Izzard, Enrico Ramirez-Ruiz and Christopher A. Tout
      \\Institute of Astronomy, Madingley Road, Cambridge, CB3 0HA}

\begin{document}
\maketitle

\label{firstpage}

\begin{abstract}

Core collapse of massive stars with a relativistic jet expulsion along
the rotation axis is a widely discussed scenario for gamma-ray burst
(GRB) production. However the nature of the stellar progenitor remains
unclear. We study the evolution of stars that may be the progenitors
of long-soft GRBs -- rotating naked helium stars presumed to have lost
their envelopes to winds or companions. Our aim is to investigate the
formation and development of single and binary systems and from this
population evaluate the rates of interesting individual species. Using
a rapid binary evolution algorithm, that enables us to model the most
complex binary systems and to explore the effect of metallicity on GRB
production, we draw the following conclusions. First we find that, if
we include an approximate treatment of angular momentum transport by
mass loss, the resulting spin rates for single stars become too low to
form a centrifugally supported disc that can drive a GRB engine --
although they do have sufficiently massive cores to form black
holes. Second massive stars in binaries result in enough angular
momentum -- due to spin-orbit tidal interactions -- to form a
centrifugally supported disc and are thus capable of supplying a
sufficient number of progenitors. This holds true even if only a small
fraction of bursts are visible to a given observer and the GRB rate is
several hundred times larger than the observed rate. Third
low-metallicity stars aid the formation of a rapidly rotating, massive
helium cores at collapse and so their evolution is likely to be
affected by the local properties of the ISM. This effect could
increase the GRB formation rate by a factor of 5--7 at
$Z=Z_\odot/200$.  Finally we quantify the effects of mass loss,
common-envelope evolution and black-hole formation and show that more
stringent constraints to many of these evolution parameters are needed
in order to draw quantitative conclusions from population synthesis
work.
\end{abstract}
\begin{keywords} gamma-rays: bursts -- stars: supernovae -- X-rays: sources \end{keywords}

\section{Introduction}

Gamma-ray bursts (GRBs) are intense flashes of gamma-rays which, for a
few seconds, light up an otherwise dark gamma-ray sky. They are
detected about once a day and outshine every other gamma-ray source in
the sky.  Major advances have been made in the last few years
including the discovery of slowly fading X-ray (Costa et al. 1997),
optical (van Paradijs et al. 1997) and radio (Frail et al. 19997)
afterglows of GRBs, the identification of host galaxies at
cosmological distances and the evidence that many of them are
associated with star forming regions and possibly supernovae (see
M\'esz\'aros 2001 for a recent review). Given the twin requirements of
enormous energy, about $10^{53}$ erg, and association with star
forming regions, the currently favoured models all involve massive,
collapsing stars and their end products, especially black holes.

Hyper-accreting black holes (0.01 to 10 $M_\odot$ s$^{-1}$) can arise
from collapsars of various types (Woosley 1993; Fryer et al. 2001);
white dwarf-black hole mergers (Frail et al. 1999; Belczynski et
al. 2002a); helium core-black hole mergers (Fryer \& Woosley 1998;
Belczynski et al. 2002a); and merging neutron stars and black holes
(Klu\'{z}niak \& Lee 1998; Ruffert \& Janka 1999; Lee \& Ramirez-Ruiz
2002). Those models that involve merging neutron stars and black holes
mostly transpire outside star forming regions and are not currently
thought to be appropriate for long GRBs (M\'esz\'aros 2001). Of the
other accretion-based models, black holes merging with both helium
cores and white dwarfs have the advantage that they could occur in
star forming regions (Belczynski et al. 2002a), produce adequate
energy, and have sufficient angular momentum to form a disc. However,
if the burst is to be beamed to less than 1\% of the sky, the white
dwarf model may provide an inadequate number of events (Fryer et
al. 1999). The helium core model requires that a hydrogen envelope
initially be present in order that the black hole (or neutron star)
experience a common envelope evolution and merge with the core. Yet
the same envelope must be absent when the burst occurs or prohibitive
baryon loading will occur (MacFadyen et al. 2001). It may be that the
merger peels off the envelope as the compact remnant goes in, but
calculations to support this hypothesis are needed. It should be noted
that the merger of a black hole with a massive helium core produces
essentially the same conditions as in ordinary collapsars, except for
a larger angular momentum. The beaming and energetics should also be
comparable.

A collapsar forms when the evolved core of a massive star collapses to
a black hole, either by fallback or because the iron core fails to
produce an outgoing shock (Woosley 1993; Paczy\'nski 1998; MacFadyen
\& Woosley 1999).  Prompt and delayed black hole formation may occur
in stars with a range of radii depending on the evolutionary state of
the massive progenitor, its metallicity and its multiplicity. The two
necessary ingredients for the collapsar are a failed or weak initial
supernova (SN) explosion that produces a black hole and sufficient
angular momentum to form a disc. Energy dissipated in the disc or the
rotation of the black hole itself is assumed to power a jet of high
Lorentz factor $\Gamma \approx 100$ (Soderberg \& Ramirez-Ruiz 2002)
which escapes the stellar progenitor along the polar axis.  

The traversal time for the relativistic jet through the hydrogen
envelope of a typical massive star ranges from hundreds to thousands
of seconds. At the time of the burst naked helium stars, which have
radii of only a few light seconds, are required if the lifetime of the
engine is not to be short when compared with the time it takes the jet
to tunnel through the star (MacFadyen \& Woosley 1999; Aloy et
al. 2000; Wheeler et al. 2000; M\'esz\'aros \& Rees 2001; \Mesz \&
Waxman 2001; Ramirez-Ruiz, Celotti \& Rees 2002; Matzner 2003). This
is aided if the stellar progenitor undergoes a Wolf-Rayet (WR) phase
characterised by a strong stellar wind that causes the star to lose
enough of its outer layers for the surface hydrogen abundance to
become minimal (MacFadyen \& Woosley 1999; Ramirez-Ruiz et
al. 2001). The radius of a WR star is then sufficiently small for the
jet to emerge before the engine ceases to operate. Break-out is
further helped if the progenitor originates in a low-metallicity
environment, when stars are smaller in radius, lose less mass and
consequently have more massive cores. Both effects are thought to
favour GRB formation (MacFadyen \& Woosley 1999; Fryer et al. 1999).

Here we present population synthesis calculations which allow us to
explore what effect initial mass, metallicity and membership in a
binary system have on GRB production. In addition to all aspects of
single-star evolution, our rapid evolution algorithm includes features
such as mass transfer, mass accretion, common-envelope evolution,
collisions, SN kicks and angular-momentum-loss mechanisms which
enables us to model even the most complex binary systems. The
binary-evolution algorithm is described in detail in
Section~2. Section~3 contains the results of binary and single
populations of core collapse SNe in solar metallicity environments. We
also quantify the effects that mass loss, common-envelope evolution,
and black-hole formation have on GRB production. The various kinds of
companion remnants that are expected when the primary massive star
collapses to a black hole are also discussed in Section~3. We present
the results of population synthesis to examine the effects of
metallicity in Section~4 and compare our model results with
observations in Section~5. Our conclusions are then given in
section~6.

\vspace{2cm}
  
\section{Stellar evolution models}

\subsection{A field guide to stellar evolution}
\begin{figure*}
\centerline{\psfig{figure=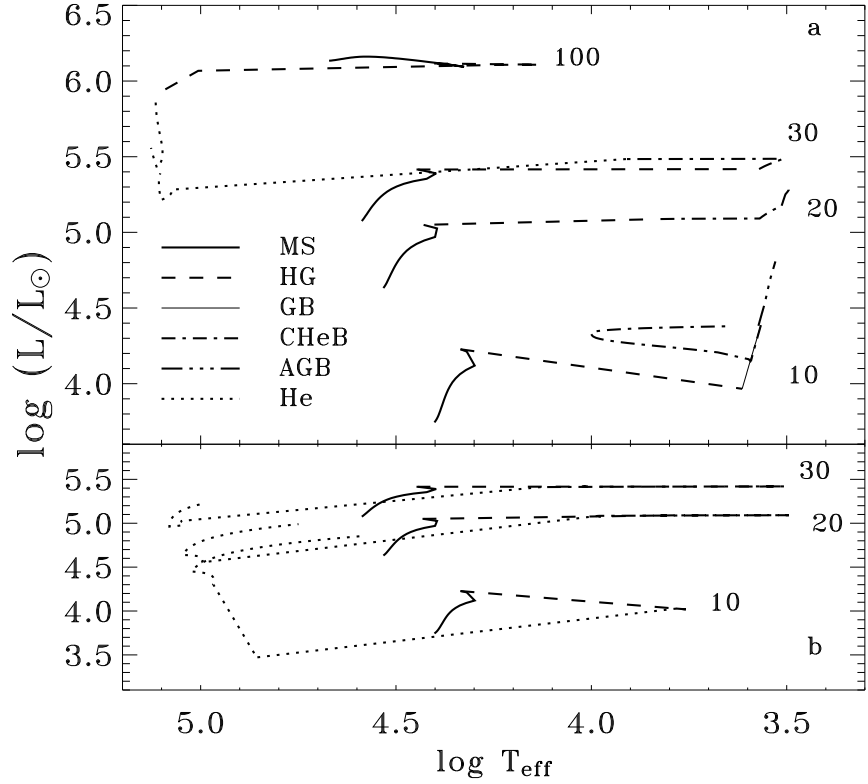,width=\textwidth}}
{\caption{Selected evolutionary tracks in the HR diagram of massive
stars ($M \ge 10 M_\odot$) with initial $Z=0.02$ composition. The
different line styles correspond to different evolutionary phases:
main sequence (MS); Hertzsprung gap (HG); first giant branch (GB);
core helium burning (CHeB); asymptotic giant branch (AGB); and naked
helium star (He). (a) Single stars evolved with mass loss by stellar
winds. (b) Binary evolution with a secondary $10 M_\odot$ star
initially orbiting the primary with a 100 days period.}
\label{fig1}}
\end{figure*}

A fundamental tool in stellar evolution is the Hertzsprung-Russell
diagram (HRD) which provides a correspondence between the observable
stellar properties -- the effective surface temperature \Teff and
luminosity $L$.  Fig. \ref{fig1} shows the evolution of a collection
of stars in the HRD from the zero-age main sequence (ZAMS), where a
star adjusts itself to nuclear and hydrostatic equilibrium, until the
end of nuclear burning.  As stars take a relatively short time to
reach the ZAMS after molecular cloud collapse, all ages are measured
from this point.  The span of a star's life, its path on the HRD and
its ultimate fate depend critically on its mass.

The immediate post-MS evolution towards the right in the HRD occurs at
nearly constant luminosity and is very rapid.  For this reason very
few stars are seen in this phase, and this region of the HRD is called
the Hertzsprung gap (HG) or the sub-giant branch. During this HG phase
the radius of the star increases greatly causing a decrease in
\Teff. As the convective envelope grows in extent the star reaches the
giant branch (GB). Eventually a point is reached on the GB where the
core temperature is high enough for stars to ignite their central
helium supply.  When core helium burning (CHeB) begins the star
descends along the GB until contraction moves the star away from the
fully convective region of the HRD and back towards the MS in what is
called a blue loop.  During CHeB, carbon and oxygen are produced in
the core.  Eventually core helium is exhausted and the star moves back
to the right in the HRD. Stars of very high mass ($M_{\rm zams} > 25 \Msun$)
reach high enough central temperatures on the HG for helium to ignite
before reaching the GB.

Evolution after the exhaustion of core helium is very similar to that
occurring after core-hydrogen exhaustion at the end of the MS -- the
convective envelope deepens again to begin what is called the
asymptotic giant branch (AGB). On the AGB the star is now a dense core
composed of carbon and oxygen surrounded by a helium burning shell
which in turn adds carbon (and some oxygen) to the degenerate core.
Initially the H-burning shell is weak so that the luminosity is supplied
by the He-burning shell. During this phase, the stellar radius grows
significantly. 

The surface gravity of the star is lowered so the surface material is
less tightly bound. Mass loss from the stellar surface becomes
significant with the rate of mass loss increasing with time.
Unfortunately our understanding of the mechanisms that cause this mass
loss is poor with possible suggestions linking it to the helium shell
flashes or to periodic envelope pulsations.  Whatever the cause, the
influence on the evolution of AGB stars is significant.  Mass loss
eventually removes all of the star's envelope so that the H-burning
shell shines through.

If the mass of the star is large enough, $M_{\rm zams} \ga 7 \Msun$
(although the exact value depends on its metallicity and mass-loss
history), the carbon-oxygen core is not very degenerate and carbon
ignites as it contracts, followed by a succession of nuclear reaction
sequences which very quickly produce an inner iron core.  Any further
reactions are endothermic and cannot contribute to the luminosity of
the star.  Photo-disintegration of iron, combined with electron
capture by protons and heavy nuclei, then removes most of the electron
degeneracy pressure supporting the core and it begins to collapse
rapidly.  When the density becomes large enough the inner core
rebounds sending a shock wave outwards through the outer layers of the
star that have remained suspended above the collapsing core.  As a
result the envelope of the star is ejected in a SN explosion so that
the AGB is truncated soon after the start of carbon burning. The
remnant in the inner core stabilises to form a neutron star (NS)
supported by neutron degeneracy pressure unless its mass is large
enough that complete collapse to a black hole occurs.

Stars with $M_{\rm zams} \ga 25 \Msun$ are severely affected by mass
loss during their entire evolution and may lose their hydrogen
envelopes, exposing nuclear processed material. If this occurs a naked
helium star is produced and such stars (or stars about to become naked
helium stars) may be WR stars. WR stars are massive objects which are
found near the MS, are losing mass at very high rates and show weak or
no hydrogen lines in their spectra. Naked helium stars can also be
produced from less massive stars in binaries as a consequence of mass
transfer (see Fig. \ref{fig1}b). Variation in composition affects the
stellar evolution timescales as well as the appearance in the HRD and
even the ultimate fate of the star. A more detailed discussion of the
various phases of evolution can be found in Hurley at el. (2000).
\vspace{-0.5cm}
\subsection{Binary evolution}
The evolution of binary stars does not differ from that of single
stars unless they get in each other's way. If the binary orbit is wide
enough the individual stars are not affected by the presence of a
companion so that standard stellar evolution theory is all that is
required to describe their development. However if the stars become
close they can interact, with consequences for the evolution. Models
for binary evolution have been implemented in the past (e.g. Whyte \&
Eggleton 1985; Pols \& Marinus 1994; Portegies Zwart \& Verbunt
1996). Here we carry out computational simulations with the
rapid-binary evolution algorithm first developed by Tout et al. (1997)
and recently updated by Hurley et al. (2002). Amongst other
improvements Hurley et al. (2002) incorporate the detailed single-star
evolution formulae of Hurley, Pols \& Tout (2000) which allow for a
wider range of stellar types than the description of stellar evolution
used by Tout et al. (1997). This requires an update of the treatment
of processes such as Roche lobe overflow, common-envelope evolution
and coalescence by collision (the reader is referred to Hurley et
al. 2002 for further details). 

As a result of Roche-lobe overflow it is possible for the binary
components to come into contact and coalesce or for the binary to
reach a common-envelope (CE) state.  The most frequent case of
common-envelope evolution involves a giant transferring mass to a
main-sequence star on a dynamical timescale.  Although the process is
difficult to model, and therefore uncertain, it is envisaged that the
secondary is not able to accept the overflowing material owing to its
relatively long thermal timescale.  The giant envelope overfills the
Roche-lobes of both stars so that the giant core and the MS star are
contained within a common-envelope.  Owing to its expansion the
envelope rotates slower than the orbit of the core and the MS star so
that friction causes them to spiral together and transfer energy to
the envelope. We assume that the cores spiral-in, transferring orbital
energy to the envelope with an efficiency ${\alpha}_{\rm CE}$, which is
necessarily a free parameter due to uncertainty in its value.  It is
probably not a constant (Reg\H{o}s \& Tout 1995) but generally
${\alpha}_{\rm CE} \approx 1$ is used.

Throughout this paper we refer to one star as the primary, mass $M_1$,
and the other as the secondary (or companion), mass $M_2$. At any time
the primary is the star filling, or closest to filling, its Roche
lobe. Numerical values of mass, luminosity and radius are in solar
units unless indicated otherwise. The algorithm provides the stellar
luminosity, radius, core mass, core radius, and angular momentum, for
each of the component stars as they evolve. A prescription for mass
loss from stellar winds is also included in the algorithm.  The
algorithm covers all the evolution phases from zero-age main sequence
(ZAMS) up to and including the remnant stages and is valid for all
masses in the range 0.1 to 100 $M_\odot$ and metallicities from
$Z=10^{-4}$ to 0.03. This rapid binary-evolution algorithm is a
natural extension of the single-star evolution algorithm. Our aim is
to evolve a population of binaries according to chosen distributions
of primary mass, secondary mass and orbital separation, in conjunction
with a realistic birth rate function and, from this population, to
calculate birth rates and expected numbers in the Galaxy for stars
that explode as type Ib/c SNe, thought to be possible GRB progenitors.

\subsection{Method}
We first set up a grid of initial binary parameters (in
the detailed single-star evolution the secondary is removed from the
system) $M_1$, $M_2$ and separation $a$ within the limits:
\begin{eqnarray}
M_{1},M_{2}: & 0.1\rightarrow 80.0\, \textrm{M}_{\odot } & \\
a: & 3.0\rightarrow 10^{4}\, \textrm{R}_{\odot } & 
\end{eqnarray}
with the $n_{\chi} = 100$ grid points of parameter ${\chi}$
logarithmically spaced,
\begin{equation}
\delta \ln {\chi} = \frac{1}{n_{\chi} - 1} 
\left[ \ln {\chi}_{\max} - \ln {\chi}_{\min} \right] 
\, . 
\end{equation}
For each set of initial parameters we evolve the binary system to an
age of $100\,$ Myr. If a binary system $j$ evolves through
a phase that is to be identified with a certain individual binary
population $i$ then the system makes a
contribution
\begin{equation}
\delta r_j = S \Phi \left( \ln M_{1j} \right) 
\varphi \left( \ln M_{2j} \right) 
\Psi \left( \ln a_j \right)\; \delta \ln M_1\; \delta \ln M_2\; \delta \ln a 
\end{equation}
to the rate $r_i$ at which that particular population is born.  This
rate depends on the star formation rate $S$, the primary mass
distribution $\Phi \left( \ln M_1 \right)$, the secondary mass
distribution $\varphi \left( \ln M_2 \right)$ and the separation
distribution $\Psi \left( \ln a \right)$. A star formation rate of
$S=7.608\, \textrm{galaxy}^{-1}\, \textrm{yr}^{-1}$ is used in this
work which corresponds to one star with $M>0.8\, \textrm{M}_{\odot} $
forming in the Galaxy every year (see Hurley et al. 2002). Such a rate
is in rough agreement with the birth rate of white dwarfs in the
Galaxy $\approx 2 \times 10^{-12}$ pc$^{-3}$ yr$^{-1}$ (Phillips
1989), noting that only stars with $M>0.8\, \textrm{M}_{\odot} $ can
possibly evolve to white dwarfs in the age of the Galaxy\footnote{The
same assumptions regarding the star formation rate have been made
previously (e.g. Iben \& Tutukov 1984; Han 1998; Hurley et al. 2002)
facilitating comparison with these results.}, and assuming an
effective Galactic volume of $\approx 5 \times 10^{11}$ pc$^{3}$. The
primary mass distribution $\xi (m)$ is the IMF of Kroupa, Tout \&
Gilmore (1993; hereafter KTG),
\begin{equation}
\xi \left( m \right) = \left\{ \begin{array} {l@{\quad}l} 
0 & m \leq m_0 \\ 
a_1 m^{-1.3} & m_0 < m \leq 0.5 \\ 
a_2 m^{-2.2} & 0.5 < m \leq 1.0 \\ 
a_2 m^{-2.7} & 1.0 < m < \infty,
\end{array} \right. 
\end{equation}
where $\xi \left( m \right) dm$ is the probability that a star has a
mass, expressed in solar units, between $m$ and $m + dm$.  The
distribution is normalized according to $\int^\infty_0 \xi (m) \, dm =
1$, so that, for $m_0=0.1$, $a_1=0.29056$ and $a_2=0.15571$. Then
$\Phi \left( \ln M_1 \right) = M_1 \, \xi \left( M_1 \right)$. If the
component masses are to be chosen independently, the secondary mass
distribution is then $\varphi \left( \ln M_2 \right) = M_2 \, \xi
\left( M_2 \right)$.  However there is observational evidence
(Eggleton, Fitchett \& Tout 1989; Mazeh et al. 1992; Goldberg \& Mazeh
1994) to support correlated masses: $\varphi \left( \ln M_2 \right) =
\frac{M_2}{M_1} = q_2$, which corresponds to a uniform distribution of
the mass-ratio $q_2$, for $0 < q_2 \leq 1$ (the flat-q
distribution). The separation distribution is taken to be $\Psi \left(
\ln a \right) = \vartheta$, a constant, between the limits 3 and $10^4
R_\odot$. Normalization gives $\vartheta = 0.12328$.

\subsection{Mass loss and rotation}
The particular mass-loss prescriptions through the various phases,
which have been found to fit the observations well, are described in detail
by Hurley et al. (2000; 2002).  On the GB phase and beyond we apply mass
loss to the envelope according to the formula of Kudritzki \& Reimers
(1978), while for the AGB we apply the formulation of Vassiliadis \&
Wood (1993).  For massive stars we model mass loss over the entire HRD
using the prescription given by Nieuwenhuijzen \& de Jager (1990),
\begin{equation}
{\dot{M}}_\roma{NJ} = \left( \frac{Z}{Z_{\odot}} \right)^{1/2} 
9.6 \times 10^{-15} R^{0.81} L^{1.24} M^{0.16} \: \Msun {\roma{yr}}^{-1} 
\end{equation}
for $L > 4000 \Lsun$, modified by the factor $Z^{1/2}$ (Kudritzki et
al.\ 1989). Numerical values of mass, luminosity and radius are in
solar units, unless otherwise specified. 

For small H-envelope mass,
$\mu < 1.0$, we include a Wolf-Rayet-like mass loss (Hamann \&
Koesterke 1998) which we have reduced to give
\begin{equation}
{\dot{M}}_\roma{WR} = 10^{-13} L^{1.5} \left( 1.0 - \mu \right) 
\: \Msun {\roma{yr}}^{-1} 
\end{equation}
where $\mu$ is given by
\begin{equation}
\label{mu_for_normal_stars}
\mu =\left( \frac{M_{\textrm{env}}}{M} \right) \min \left[5.0,\max
\left\{1.2, \left( \frac{L}{L_{0}} \right)^{\kappa}\right\}\right].
\end{equation}
Here $L_{0}=7.0\times 10^{4}\, \textrm{L}_{\odot }$ and $\kappa =-0.5$
(Hurley et al. 2000). The reduction in ${\dot{M}}_\roma{WR}$ is
necessary in order to produce sufficient black holes to match the number
observed in binaries (see Hurley et al. 2000).  

As we plan to use the
evolution routines to investigate GRB production, it is desirable to
follow the stars' angular momentum. To do this we must start each star
with a realistic spin on the ZAMS.  A reasonable fit to the
${\bar{v}}_\roma{rot}$ MS data of Lang (1992) is given by 
\beq
{\bar{v}}_\roma{rot} \left( M \right) = \frac{330 M^{3.3}}{15.0 +
M^{3.45}} \: \roma{km} \, {\roma{s}}^{-1} \eeq so that \beq \Omega =
45.35 \frac{{\bar{v}}_\roma{rot}}{\RZAMS} \quad \roma{yr^{-1}} \, .
\eeq 
The angular momentum is then given by
\begin{equation} 
J_\roma{spin} = I \Omega = k M R^2 \Omega 
\end{equation}
where the constant $k$ depends on the internal structure, e.g. $k =
2/5$ for a solid sphere and $k = 2/3$ for a spherical shell.  In
actual fact we find the angular momentum by splitting the star into
two parts, consisting of the core (with mass \Mc) and the envelope
(with mass \Menv), so that \beq J_\roma{spin} = \left( k_2 \left( M -
\Mc \right) R^2 + k_3 \Mc {R_\roma{c}}^2 \right) \Omega \eeq where
$k_2 = 0.1$, based on detailed giant models which reveal $k = 0.1
\Menv / M$, and $k_3 = 0.21$ for an $n = 3/2$ polytrope such as a
white dwarf, neutron star or dense convective core.  This works well
for post-MS stars which have developed a dense core whose rotation is
likely to have decoupled from the envelope while also representing the
near uniform rotation of homogeneous MS stars which have no core. When
the star loses mass its stellar wind carries off angular momentum at a
rate given by \beq \dot{J} = k \dot{M} h \eeq where $h = R^2 \Omega$.
Thus \beq J_\roma{spin} (t+\Delta t) = J_\roma{spin} (t) - \frac{2}{3}
\Delta M R^2 \Omega \eeq when the star loses an amount of mass $\Delta
M$, where we take $k = 2/3$ because we assume that all the mass is
lost uniformly at the surface of the star ie. from a spherical shell.

We also include magnetic braking for stars that have appreciable
convective envelopes 
\beq 
{\dot{J}}_\roma{mb} = 5.83 \times
10^{-16} \frac{\Menv}{M} \left( R \Omega \right)^{3} \: \Msun
{\roma{R}}^2_{\odot} {\roma{yr}}^{-2},
\eeq
with $\Omega$ in units of years. Following Rappaport et al.\ (1983) we
don't allow magnetic braking for fully convective stars, $M <
0.35M_\odot$ but this restriction does not affect our results here.

\subsection{Core collapse progenitors}

Core-collapse SNe spectra may contain hydrogen absorption lines (type
II) or show very weak or no hydrogen lines at all (type Ib/c) thought
to indicate the lack of a hydrogen envelope around the imploding
star. As in Hurley et al. (2002), we define a type Ib/c SN as one that
produces a neutron star or black hole from a primary, naked helium
star. All other SN progenitors that produce a neutron star or black
hole are considered to be type II. The carbon-oxygen core mass $M_{\rm
CO}$ of a SN~Ib/c progenitor has a maximum value of $\max [M_{\rm Ch},
0.773M_{\rm He}-0.35\, \rm{M}_\odot]$ (Hurley 2000; hereafter H00),
where $M_{\rm He}$ is the initial mass of the helium star and $M_{\rm
Ch}=1.44\, \rm{M}_{\odot }$ is the Chandrasekhar mass. For masses
above this, a SN explosion is assumed to take place. The expression
for the maximum mass of the core of a SN~II progenitor is similar with
$M_{\rm He}$ replaced by the mass of the carbon-oxygen core at the
base of the asymptotic giant branch. If \beq M_{\rm CO}>7.0\,
\textrm{M}_{\odot}
\label{mco}
\eeq we assume a black hole forms (Woosley 1993). The mass of the
neutron star (NS) or black hole (BH) formed is then given by (H00)
\beq M_{\rm NS/BH}=1.17+0.09M_{\rm CO}\, \rm{M}_{\odot}
\label{BH}
\eeq corresponding to neutron star baryonic masses in the range $1.3\,
\textrm{M}_{\odot }$ to $1.8\, \textrm{M}_{\odot }$ and a minimum
black hole mass of $1.8\, \textrm{M}_{\odot }$.
  
A Wolf-Rayet stellar wind $\dot{M}_{\textrm{WR}}$ is used in the
stellar models when the H-envelope mass $M_{\textrm{env}}$ is small
$\mu <1.0$. Stars with $\mu >1.0$ are excluded from the
Wolf-Rayet population. 

We apply the condition that a star which is to be considered as a GRB
progenitor must: (i) go through a Wolf-Rayet phase, as defined above,
before ending its life as a type Ib/c SN; (ii) have a sufficiently
massive core to form a black hole; and (iii) have enough angular
momentum at the time of collapse to allow the formation of a disc. To
decide whether a centrifugally supported accretion disc can form
around a central black hole we compare the angular momentum of the
progenitor star to that which a test particle would require at the
last stable orbit (LSO) around a black hole (see e.g. Heger \& Woosley
2002). In the case of a non-rotating, spherically symmetric black hole
the last stable orbit has a radius $r_{\rm LSO} =6GM_{\rm
BH}/c^2$. The condition that the matter does not spiral into the black
hole is simply that the rotational energy of the matter should be
greater or equal to half the gravitational potential energy: ${1 \over
2}J^2/I \le {1 \over 2} GM_{\rm BH}m/r_{\rm LSO}$, where $J$ is the
angular momentum of the element mass $m$ and $I=mr_{\rm LSO}^2$ is its
moment of inertia about the hole in the LSO. In other words, $j > G
M_{\rm BH} r_{\rm LSO}^{1/2}$, where $j$ is the specific angular
momentum $j=J/m$. The inner boundary condition is therefore that the
angular momentum with which the material arrives at the stable orbit
should be greater than or equal to this critical value.

\section{Progenitors of Core Collapse SN\lowercase{e} and GRB\lowercase{s}}

\subsection{Supernovae}
The effects of close binary evolution are observed in many systems,
such as cataclysmic variables, X-ray binaries and Algols and by the
presence of stars such as blue stragglers which cannot be explained by
single star evolution.  The fraction of star systems containing at
least two gravitationally bound stars, i.e. the binary fraction, is at
least 50\% (see Larson 2001 for a recent review). In the case of
Wolf-Rayet stars it is perhaps as high as 80\%.  Some uncertainty still
exists due to small number statistics but multiplicity seems a crucial
element of massive star evolution.

While many of the processes involved are not
understood in detail we do have a qualitative picture of how binaries
evolve and can hope to construct a model that correctly follows them
through the various phases of evolution. In order to conduct
statistical studies of complete binary populations, i.e. population
synthesis, such a model must be able to produce any type of binary
that is observed in enough detail. 

For most of the individual binary populations, observational birth
rates or numbers in the Galaxy are uncertain because of selection
effects involved when undertaking surveys.  However enough data exist
overall to enable a meaningful comparison with the
results.  Cappellaro~(2001) combined the results of five independent SN
searches to obtain a sample of 137 SNe from which he derived the
following birth rates for the Galaxy:
\begin{eqnarray*} 
2.4 \pm 1 \times 10^{-3} & {\rm yr}^{-1} & \mbox{  SNIb/c}\\
14.9 \pm 6 \times 10^{-3} & {\rm yr}^{-1} & \mbox{  SNII}. \\ 
\end{eqnarray*} 
The values reported above assume $H_0=65$ km s$^{-1}$ Mpc$^{-1}$, and
that the Galaxy has an average SN rate its morphological type
(here assumed Sb-Sbc) and luminosity ($2.3 \times 10^{10}
L_{B,\odot}$; see Cappellaro et al. 1999).  Our predicted formation
rates for SNe~Ib/c and SNe~II in single and binary stellar populations
are shown in Table 1. Binary models have been evolved according to
chosen distributions of primary (KTG distribution) and secondary (KTG
or q-flat distribution) masses. Throughout this paper we refer to
these distributions as KTG-KTG or KTG-q, respectively. The single star
results are consistent with the observed rates, as are the binary type
II SNe at $Z = 0.02$.  A discrepancy lies in the birth rates of type
Ib/c SNe because the theoretical rates in the KTG-q case are high but
not for the KTG-KTG case which are in good agreement. This can be
reconciled by assuming a binary fraction smaller than unity (Hurley et
al. 2002).

The SN~II rate is a fairly robust number but the SN~Ib/c rate is
sensitive to many of the assumptions underlying the model (Hurley et
al. 2002). In binary models the majority of type Ib/c SNe come from
naked helium stars formed from relatively low-mass progenitors ($10-20
\Msun$) stripped by the interaction, while single SNe Ib/c arise
mainly from more massive stars $> 24 \Msun$ which become Wolf-Rayet
stars prior to explosion (see Fig. \ref{fig2}a).  In a model
containing only single stars with $Z=0.02$ the SN~Ib/c rate is reduced
to $2\times 10^{-3} \, {\rm yr}^{-1}$ which would be consistent with
observations.  These rates are sensitive to the assumed mass-loss rate
for red supergiants (see Hurley et al. 2000), as well as to our
assumption that black hole formation ignites a SN explosion, both of
which are very uncertain.  If mass-loss rates are lower or if black
holes form without a SN, the SNIb/c rate from single stars would be
lower.

\subsection{Gamma-ray bursts}

GRBs are thought to be produced when the evolved core of a massive
star collapses to a black hole and the remaining star has too much
angular momentum to fall in directly.  Using the rapid binary code we
are able to generate a series of large single and binary populations
and evaluate the formation rate of interesting core collapse
species. Rotating naked helium stars, presumed to have lost their
envelopes to winds or companions, are evolved from the ZAMS ignition
upto the formation of the final CO core.

\begin{figure*}
\centerline{\psfig{figure=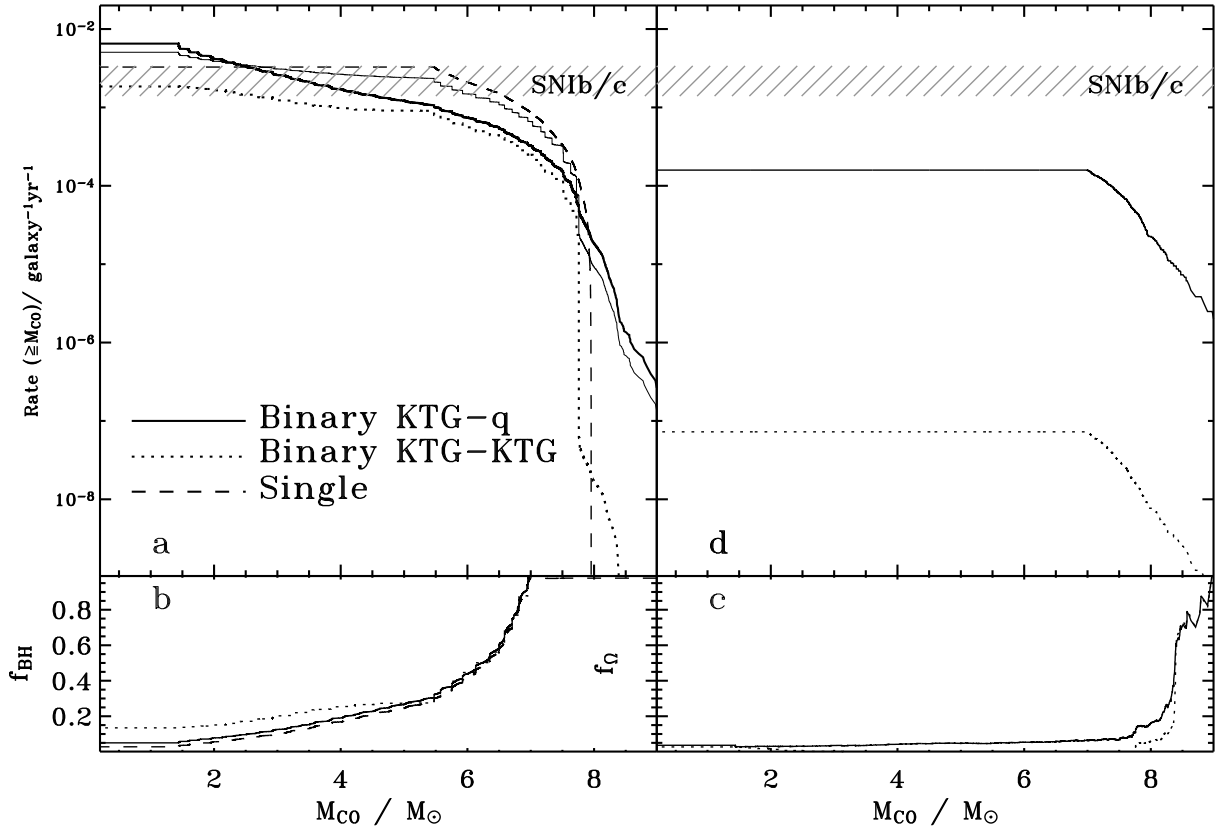,width=\textwidth}} {\caption{Formation
rates of interesting core collapse species.  (a) Cumulative rate
(above a threshold core mass $M_{\rm CO}$) for $Z=Z_\odot$ stars that
explode as type Ib/c SNe. The shaded region represents the 1$\sigma$
uncertainties on the birth rates of type Ib/c SNe as derived by
Cappellaro (2001). The birth rates reported above assume $H_0=65$ km
s$^{-1}$ Mpc$^{-1}$, and that the Galaxy has an average SN rate for
its morphological type (here assumed Sb-Sbc) and luminosity ($2.3
\times 10^{10} L_{B,\odot}$). The thin solid line assumes a 50\%
binary fraction (KTG-q case). (b) Fraction of SN Ib/c ($\ge M_{\rm
CO}$) that form a central black hole after exploding. (c) Fraction of
SN Ib/c ($\ge M_{\rm CO}$) that end their lives with sufficient
angular momentum to form a centrifugally supported disc. (d)
Cumulative rate ($\ge M_{\rm CO}$) of stars -- thought to be GRB
progenitors -- that have both a massive enough core to form a black
hole and adequate rotation rate at the time of collapse to allow the
formation of a disc.}
\label{fig2}}
\end{figure*}

The cumulative rate (above a threshold core mass, $M_{\rm CO}$) for
stars of $Z=Z_\odot$ that explode as type Ib/c SNe is shown in
Fig. \ref{fig2}a. A fraction, $f_{\rm BH}$, of the above stars will
form a central black hole after exploding (see Fig. \ref{fig2}b).  By
including angular momentum transport by non-magnetic processes, such
as mass loss, mass transfer and mass accretion, we are able to roughly
estimate the fraction $f_{\Omega}$ of SN Ib/c progenitors that end
their lives with sufficient angular momentum to form a centrifugally
supported disc (see Fig. \ref{fig2}c).

\begin{figure*}
\centerline{\psfig{figure=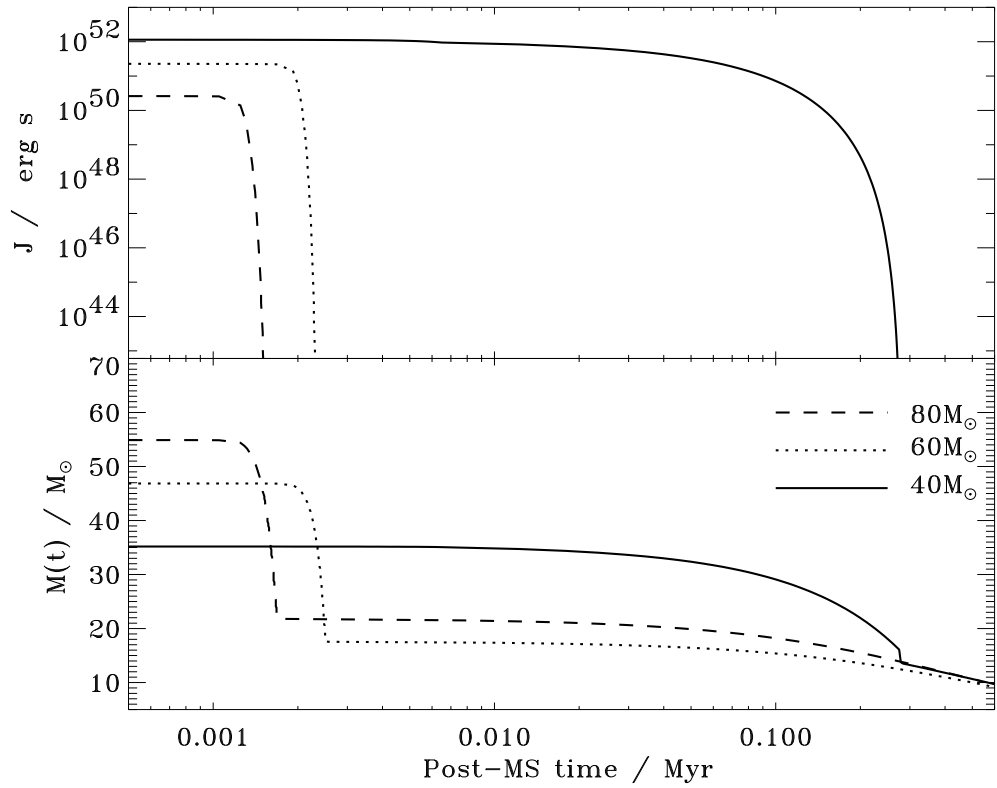,width=\textwidth}}
{\caption{Single stellar evolution after core hydrogen burning for
$Z=Z_\odot$ and initial masses 40, 60, 80 $M_\odot$. Both the mass
and angular momentum are shown for the post-MS phase as a function of time.}
\label{fig3}}
\end{figure*}

High angular momentum is a common requirement in all current GRB
models that involve massive stars, including those that use a pulsar
power source. Yet the actual angular momentum in a presupernova star
is unknown. Without magnetic fields, and with our approximate
treatment of angular momentum transport, we find that a naked helium
star cannot retain enough angular momentum to form a centrifugally
supported disc around the collapsed object, as required by the
collapsar model of GRBs ($f_{\Omega} \approx 0$ for single stars; see
Fig. \ref{fig2}c). The main reason is that during the early phase of
WR evolution, the progenitor star spins down significantly as it loses
a large fraction of its mass quite rapidly (Fig. \ref{fig3}). This is
even more so if magnetic fields couple the core effectively to the
envelope -- the magnetic interaction of the rapidly rotating helium
core with its stationary envelope during the red supergiant phase will
halt the core, making it unfit as a collapsar progenitor (Spruit \&
Phinney 1998; Spruit 2002; Heger \& Woosley 2002; but see Livio \&
Pringle 1998).  Either the description of the stellar model is
inaccurate or some route other than single star evolution must be
involved in making GRBs. The natural alternative is a close binary.

\begin{figure*}
\centerline{\psfig{file=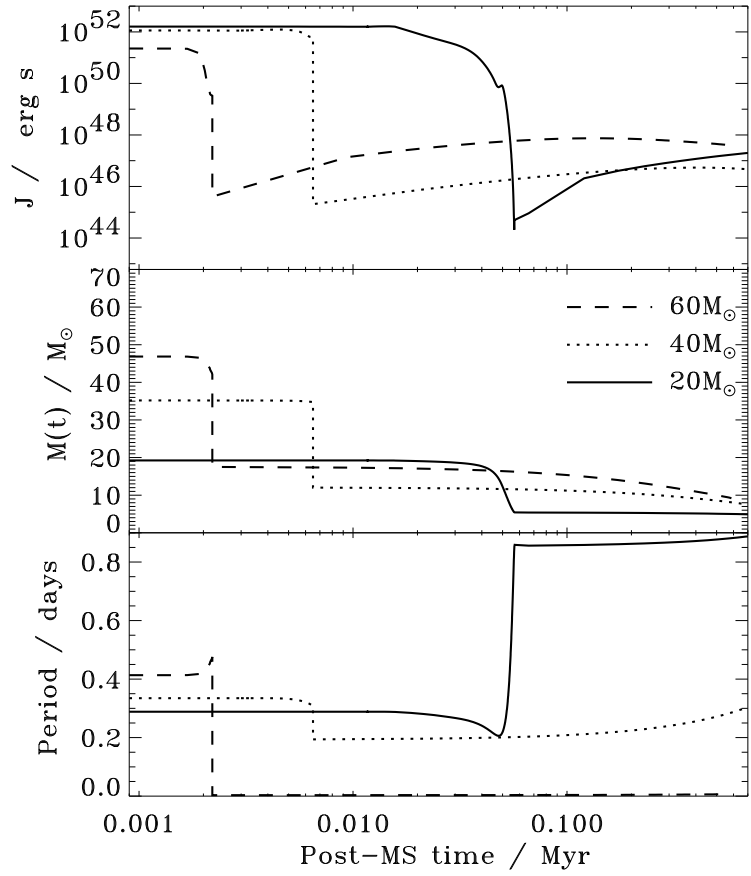,width=\textwidth}} {\caption{The
evolution of the mass and angular momentum of various primary stars
after core hydrogen burning for $Z=Z_\odot$ and initial masses 20, 40,
60 $M_\odot$ with  a binary companion with initial P=100 days and
$M_2=10M_\odot$.}
\label{fig4}}
\end{figure*}

\begin{figure*}
\centerline{\psfig{file=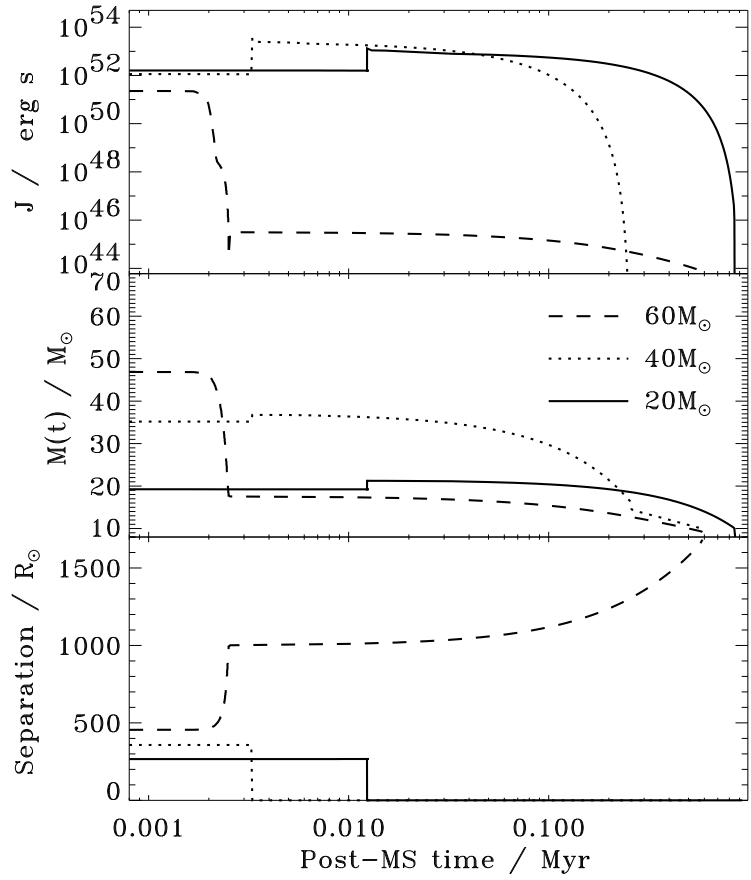,width=\textwidth}} {\caption{Similar to
Fig. \ref{fig4} but with $M_2=3M_\odot$.}
\label{fig5}}
\end{figure*}

If either star fills its Roche lobe, then gas flows from the outer
layers of the star through the inner Lagrangian point that connects
the two Roche lobes. Some or all of this gas may be captured by the
companion star so that mass transfer occurs and, as a result, the
subsequent evolution of both stars takes a different course from that
of isolated stars with important consequences for their orbit and spin
(Fig. \ref{fig4}). If the hydrogen envelope is removed too early (or
the two stars coalescence) mass loss from the WR star also reduces the
angular momentum (Fig. \ref{fig5}). The effect that the companion mass
has on the outcome of binary evolution is illustrated by comparing the
evolution of the primary star in Figs. \ref{fig4} and \ref{fig5}.\\

Finally, the expected birth rates of stars in the Galaxy that have
both a massive enough core to form a black hole and an adequate
rotation rate at the time of collapse to allow the formation of a disc
are shown in Fig. \ref{fig2}d. These core collapse species may be the
progenitors of the common (long-soft) GRBs.
\vspace{-0.5cm}
\subsection{Sensitivity to model parameters}
\begin{figure*}
\centerline{\psfig{figure=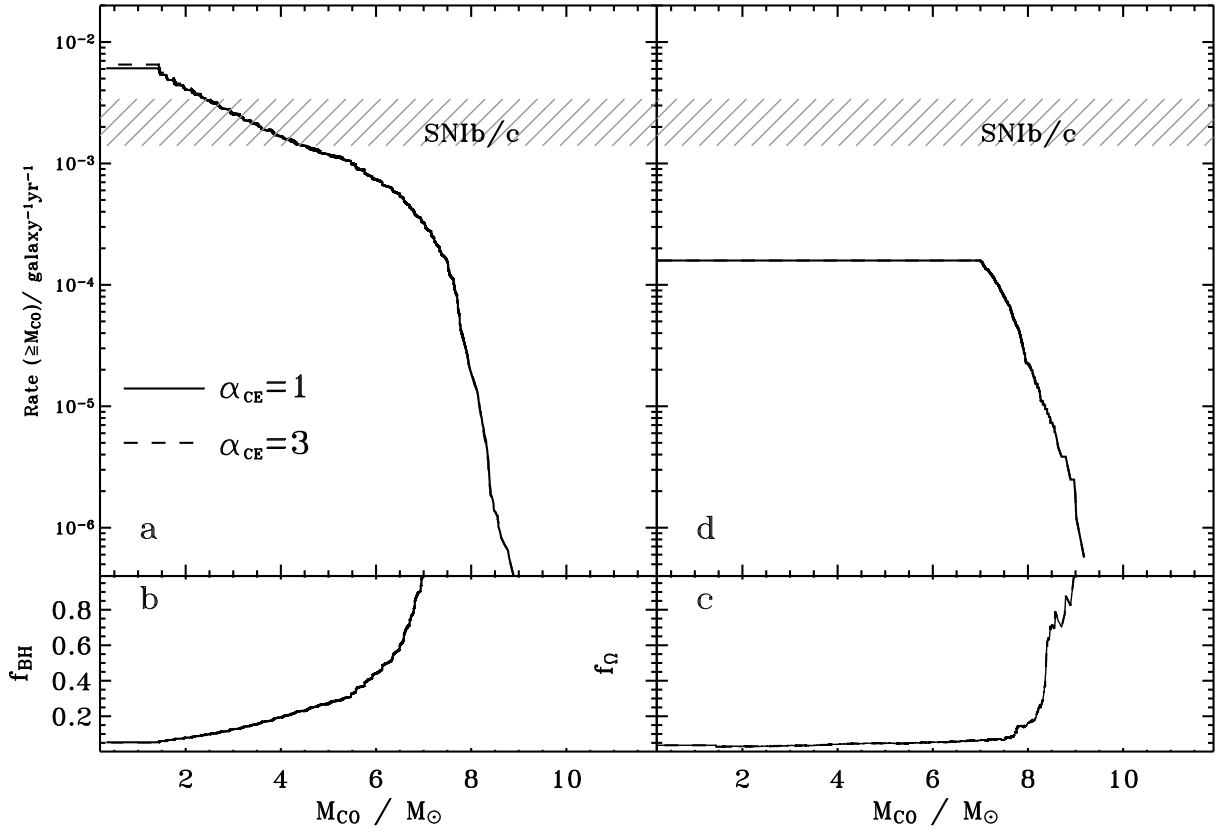,width=\textwidth}} {\caption{Similar
to Fig. \ref{fig2} but with ${\alpha}_{\rm CE}=3$ using an initial
KTG-q distribution of binaries.}
\label{fig6}}
\end{figure*}

To give an idea of how sensitive the final state of the system is to
changes in the physical parameters that govern the evolution, we
reconsider the example illustrated in Fig. \ref{fig2}. If the
common-envelope efficiency of the component stars is taken to be
${\alpha}_{\rm CE} = 3$, rather than ${\alpha}_{\rm CE} = 1$, less
energy is required to drive off the common envelope. The result is
similar to a reduction in the envelope binding energy and so it is
less likely that the process ends in coalescence of the cores. This
increment in the common-envelope efficiency parameter slightly
increases the GRB production rate (see Fig. \ref{fig6}). Wider systems
are not brought within an interaction distance after the
common-envelope phase. Thus systems that do interact when
${\alpha}_{\rm CE} = 1$ do not do so when ${\alpha}_{\rm CE} =
3$. However, the net increase is mainly because closer systems that
would coalesce when ${\alpha}_{\rm CE} = 1$ now remain detached.

At first sight this may seem an unphysical model given the definition
but, as discussed by Iben \& Livio (1993), an increase in
${\alpha}_{\rm CE}$ can be possible if additional energy sources other
than the orbital energy are involved.  Processes with potential to
supply such energy include enhanced nuclear burning in shell burning
zones of giants, nuclear burning on the surface of a degenerate
secondary, dynamo generation of magnetic fields and recombination of
the hydrogen and helium ionization zones in giants.  Possibly the
common envelope absorbs ordinary nuclear energy in the process of
swelling up, but this should occur on a thermal timescale.
Unfortunately, the theoretical determination of reliable values for
${\alpha}_{\rm CE}$ has proven difficult owing to a lack of
understanding of the processes involved and our inability to model
them.

\begin{figure*}
\centerline{\psfig{figure=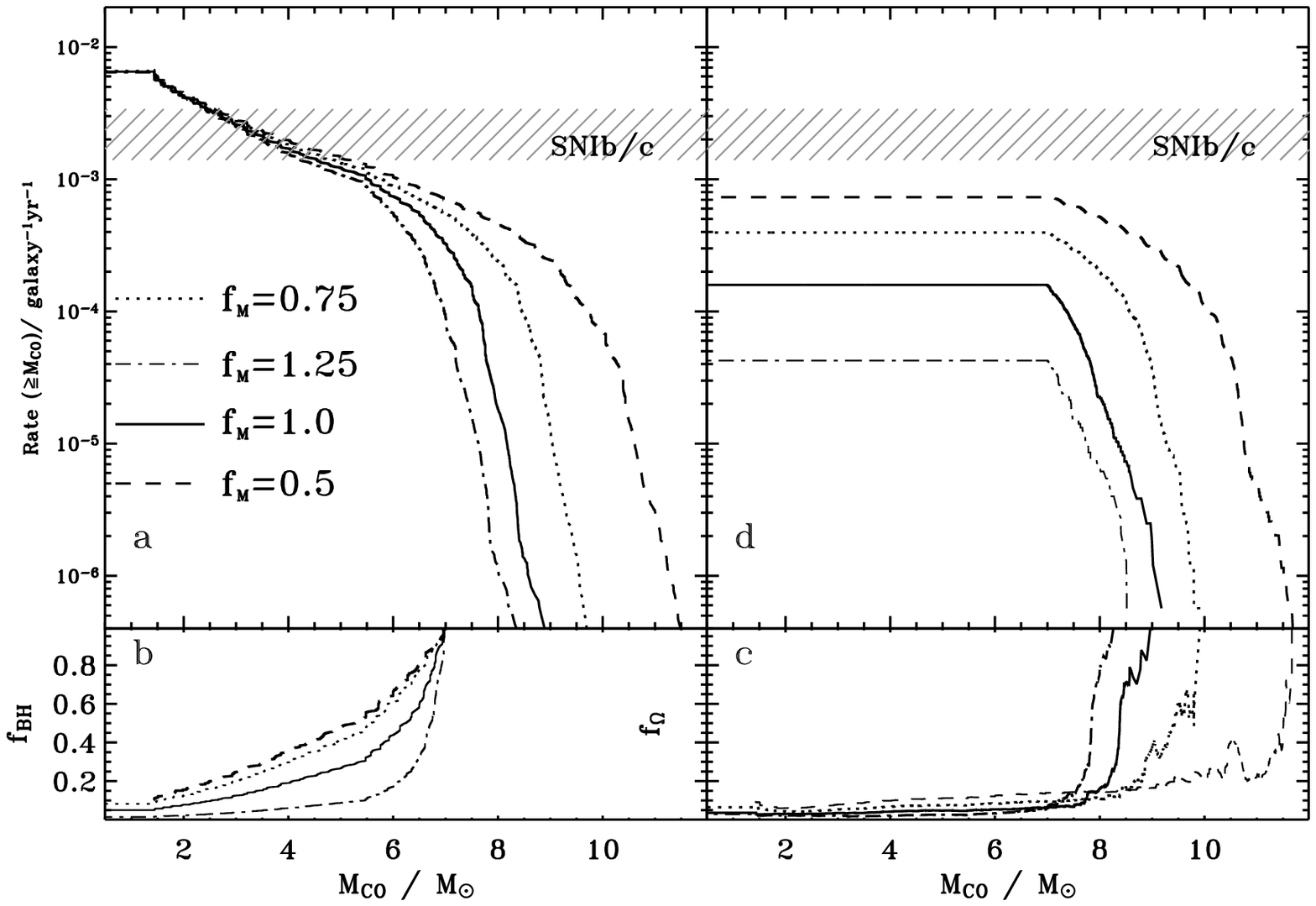,width=\textwidth}}
{\caption{Similar to Fig. \ref{fig2} for various $f_M$ using an
initial KTG-q distribution of binaries.}
\label{fig7}}
\end{figure*}

An increase (or decrease) of the mass-loss rate has profound
consequences for the WR population. At a given metallicity, the
minimum initial mass for the formation of a WR star (lower for higher
$\dot{M}$), the duration of the WR stage (greater for higher
$\dot{M}$), the times spent in the different WR subtypes and the
surface composition during these phases are all very sensitive to the
mass-loss rates (Maeder 1991). There is no doubt that major changes to
the outcome result if $\dot{M}_{WR}$ is altered by even a small
fraction $f_M$. Inspection of Fig. \ref{fig7} reveals that as the
mass-loss is reduced, the threshold for the removal of the hydrogen
envelope by stellar winds is raised and the loss of angular momentum
is inhibited. This increases the mass of the CO core and favours black hole
formation. Conversely, increasing the mass-loss rate promotes the loss
of angular momentum and inhibits black hole formation (see
Fig. \ref{fig7}).

\begin{figure*}
\centerline{\psfig{file=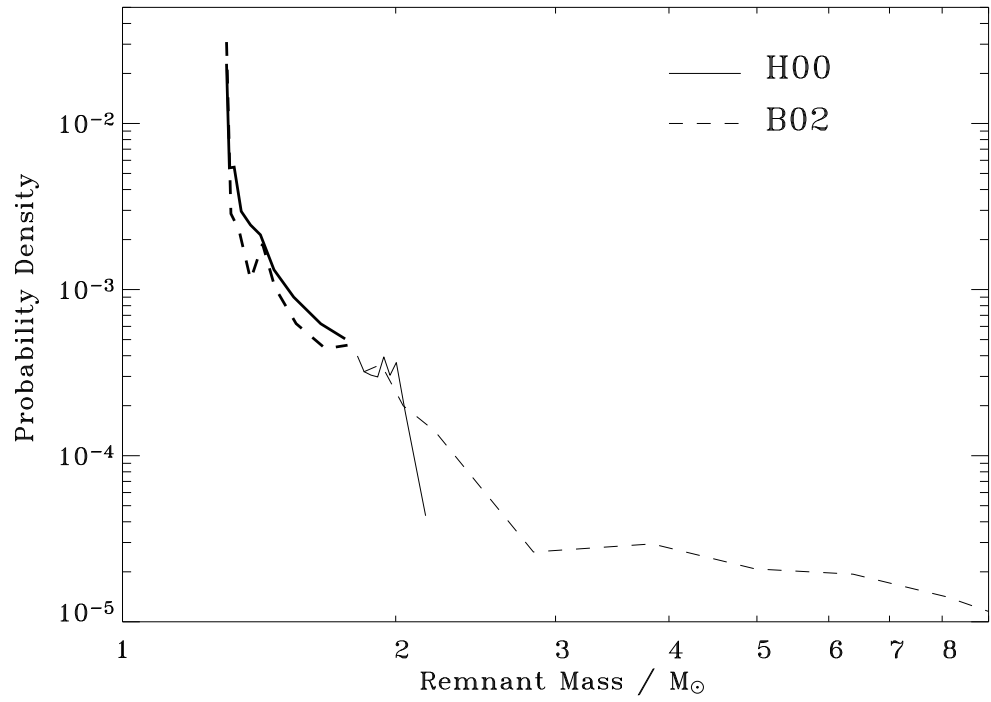,width=\textwidth}} {\caption{The
probability distribution of final remnant masses for an initial KTG-q
distribution of binaries and for two different prescriptions for NS
(thick lines) and BH (thin lines) masses. See text for details.}
\label{fig8}} 
\end{figure*}

It should be noted that the choice of the initial mass above which
stars become black holes rather than neutron stars is not well
constrained.  The low-mass X-ray binary (LMXB) catalogue compiled by
van Paradijs (1995) lists about 80 bright persistent sources in the
Galaxy but makes no distinction as to whether the compact star is a
neutron star or black hole. While this number is in fair agreement
(H00) with the birth rates derived with the prescription for the
remnant's mass given in equation (\ref{BH}), the differences
introduced by this choice in the overall evolution are small when
compared with those resulting from varying the (very uncertain)
initial distributions. The birth rate of persistent neutron star LMXBs
derived by means of such a prescription is similar to the rate found
by Portegies Zwart et al. (1996) in their standard model that includes
velocity kicks. Measurements of black hole masses in binaries,
although still highly uncertain, yield values between 3 and $20
M_\odot$ (Orosz et al. 2001; McClintock et al. 2001; Froning \&
Robinson 2001; Wagner et al. 2001; Table 1 of Fryer \& Kalogera 2001
and references therein). For progenitors of $40 \le M_{\rm
zams}/M_\odot \le 100$, equation (\ref{BH}) gives final remnant masses
in the range $1.8-3.0 M_\odot$ (Fig. \ref{fig8}). If the amount of
mass that a black hole can accrete is limited by the Eddington limit
(Cameron \& Mock 1967) then a black hole is unlikely to accrete enough
material to take its mass above $5 M_\odot$ and this model is
inconsistent with the observations. There is, however, some
uncertainty as to whether the Eddington limit should actually be
applied because the energy generated in excess of the limit might be
removed from the system in a strong wind or asymmetrically through a
disc. Super-Eddington accretion rates may be crucial in allowing black
holes in binaries to gain a significant amount of mass.

\begin{figure*}
\centerline{\psfig{figure=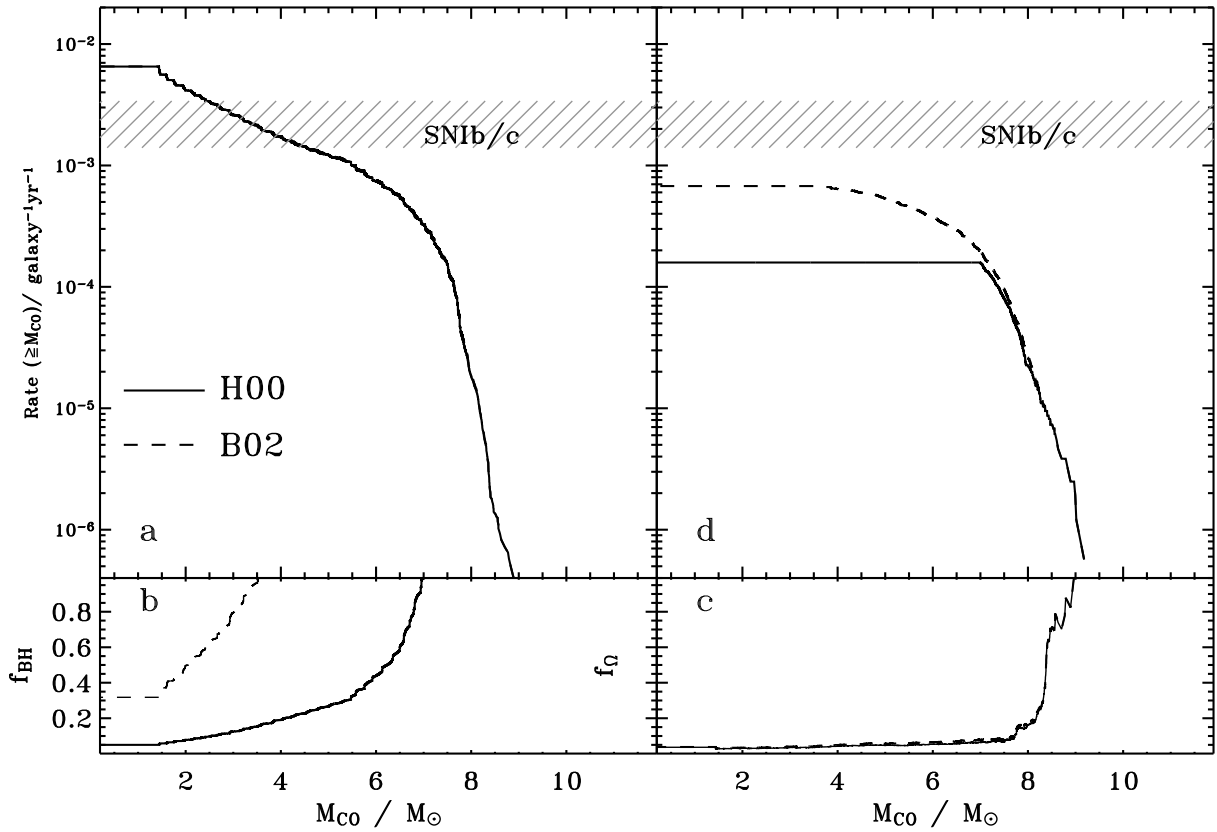,width=\textwidth}} {\caption{Similar to
Fig. \ref{fig2} but using the B02 prescription for NS and BH masses rather than
the H00 formulae. An initial KTG-q distribution of binaries is
adopted.}
\label{fig9}}
\end{figure*}

Imposing the Eddington limit significantly reduces our ability to grow
black hole masses to the observed values and, for this reason, in
addition to our somewhat arbitrary prescription (see equation
\ref{BH}) for neutron star and black hole masses, we also use the
formalism developed by Belczynski, Kalogera \& Bulik (2002b; hereafter
B02). Based on the Hurley et al. (2000) formulae, they calculate the
final CO core of a star and then use stellar models of Woosley (1986)
to obtain a final FeNi core mass as a function of the CO core mass
(see their equation [1]). To estimate the mass of the remnant formed,
B02 follow the results of hydrodynamical core-collapse calculations
(Fryer et al. 1999; Fryer \& Kalogera 2001).  Fig. \ref{fig9} compares
the black hole birth rates derived using the B02 formalism with those
based on equation (\ref{BH}). The birth rate of black holes shows an
increase for stars with low CO core masses. Comparing the birth rates
of GRBs in both models reveals a variation by a factor of $5$. The
relevance of the relative distribution of neutron stars and black
holes on the birth rates of GRBs is further complicated by the fact
that a rapidly rotating neutron stars with an ultrahigh magnetic field
could, in principle, also serve as a trigger (Usov 1992; Thompson
1994).

These examples serve to demonstrate the sensitivity of binary
evolution to the choice of model parameters.

\subsection{Stellar companions}

One of the most important questions relating to WR stars is whether
they are all members of binary systems or rather do some truly single
objects exist. Possibly the majority of stars are members of binary or
multiple systems but most are sufficiently far from their companions
that their evolution proceeds essentially as if they were single
stars. However there is an important minority of stars which are close
enough that their evolution is dramatically changed by the presence of
a companion. The most dramatic effects occur when one of the stars in
the system is a very compact object. Here we investigate the expected
binary systems with black hole primaries which may be the sources of
GRBs earlier in their evolution.

\begin{figure*}
\centerline{\psfig{file=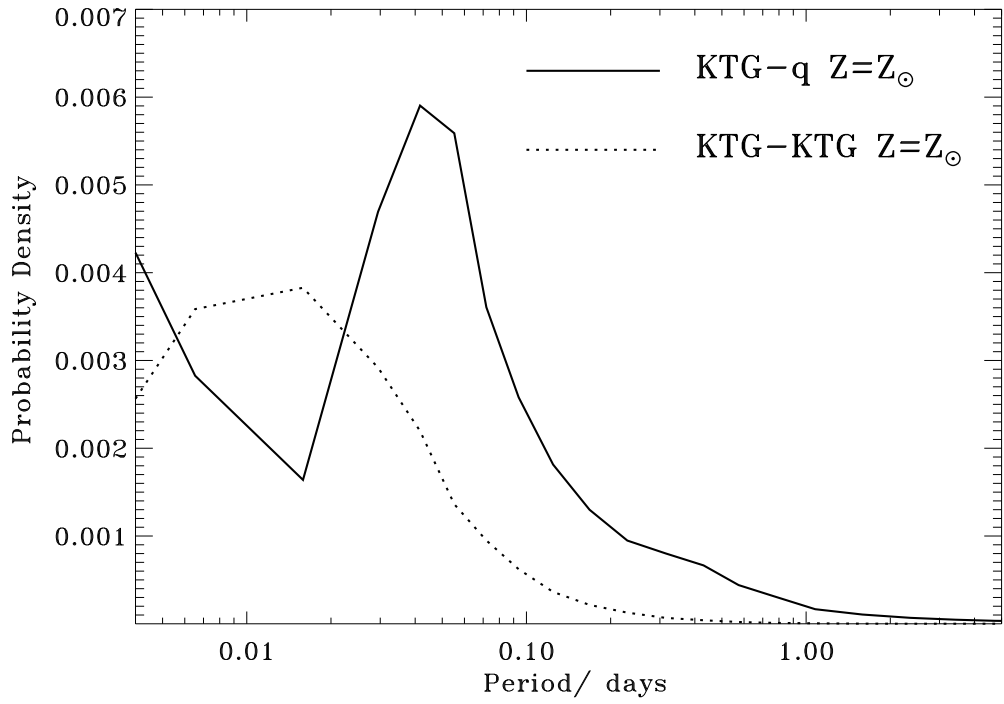,width=\textwidth}} {\caption{The
probability distribution of binary period at the time when one of
the stars explodes. The initial explosion then produces a BH and
sufficient angular momentum to form a disc.}
\label{fig10}} 
\end{figure*}

Table 2 shows the various types of companions to the GRB progenitor
star at the time of collapse. For both secondary mass distributions
the vast majority of companions are, without surprise, MS stars and
thus likely donors of soft X-ray binaries (Lee et al. 2002). The
typical orbital periods for these binaries at the time of explosion
are in the 0.01 -- 1 days range (Fig. \ref{fig10}). Most companions,
being MS stars, have small mass-loss rates of $10^{-8}M_\odot {\rm
yr}^{-1}$ or less and are thus unlikely to significantly enhance the
density around the GRB progenitor. There is a small, interesting
sub-class of binaries in which the companion star is a collapsed
object (about 1\% for the KTG-q distribution). In this case, the GRB,
which is likely caused by the relativistic jet, expelled along the
rotation axis of the collapsing stellar core, propagates within a SN
remnant of about 1 Myr in age.

\section{The role of metallicity}

The assumption that all stars within the population are born with the
same composition is somewhat naive: nucleosynthesis in successive
generations of stars enriches the gas from which they form as the
Galaxy evolves.  Metallicity $Z$ influences the stellar evolution of
massive stars mainly through bound-free and line opacities in the
outer layers of massive stars owing to their influence on stellar wind
mass loss rates. The numbers of WR stars in galaxies of different
metallicities are an important test of the stellar models at various
$Z$ and of the values of the final stellar masses. The theoretical
predictions of the ratios of WR/O-stars, WC/WR and WC/WN are in
good agreement with observations of the Local Group (Maeder 1991),
which support the following dependence of the mass-loss rates on
metallicity: $(\dot{ M_z}/\dot{M}_{\odot}) = (Z/Z_\odot)^{0.5}$ (de
Jager et al. 1988).

\begin{figure*}
\centerline{\psfig{figure=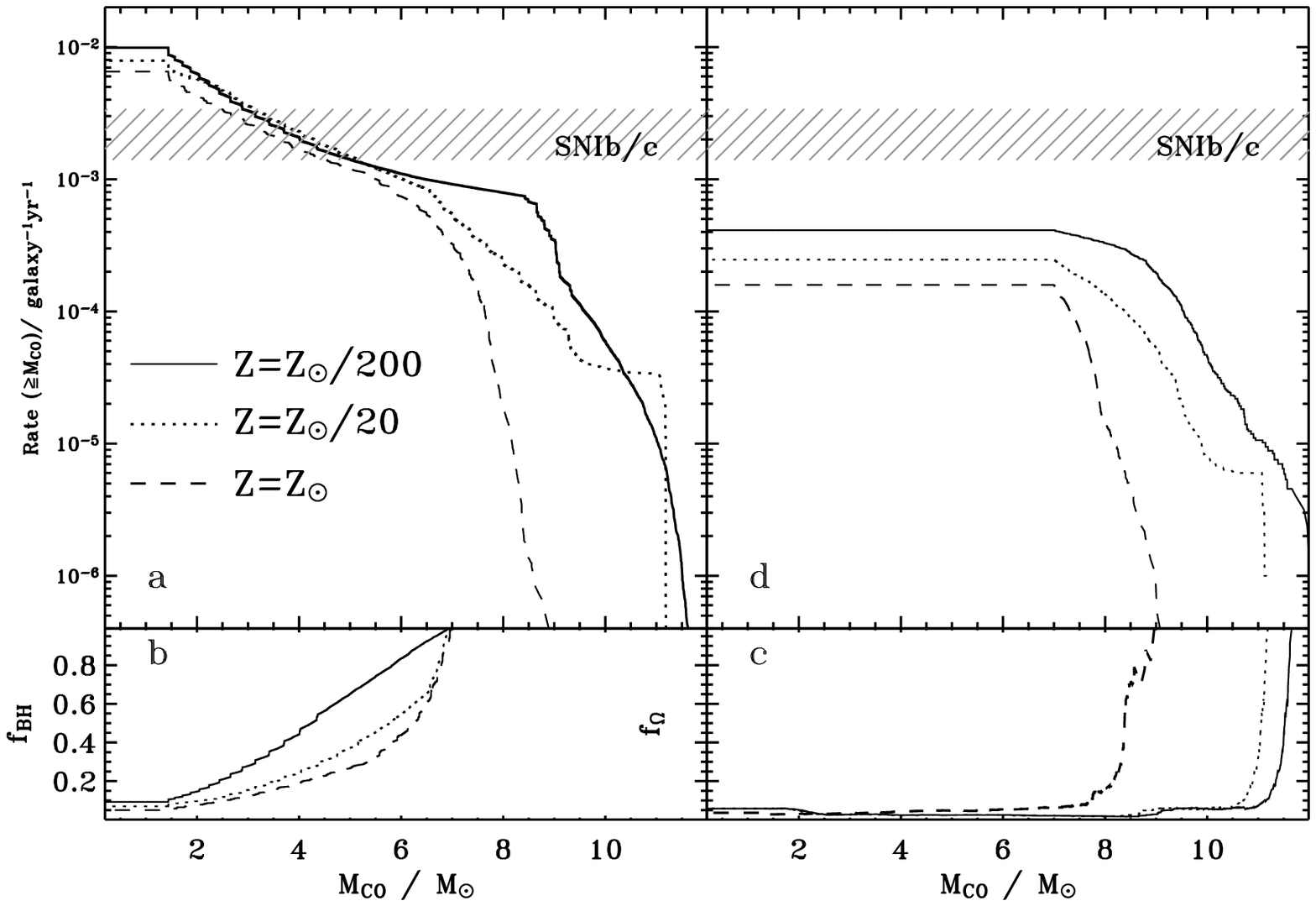,width=\textwidth}} {\caption{Similar
to Fig. \ref{fig2} for various $Z$ using an initial KTG-q distribution
of binaries.}
\label{fig11}}
\end{figure*}

\begin{figure*}
\centerline{\psfig{figure=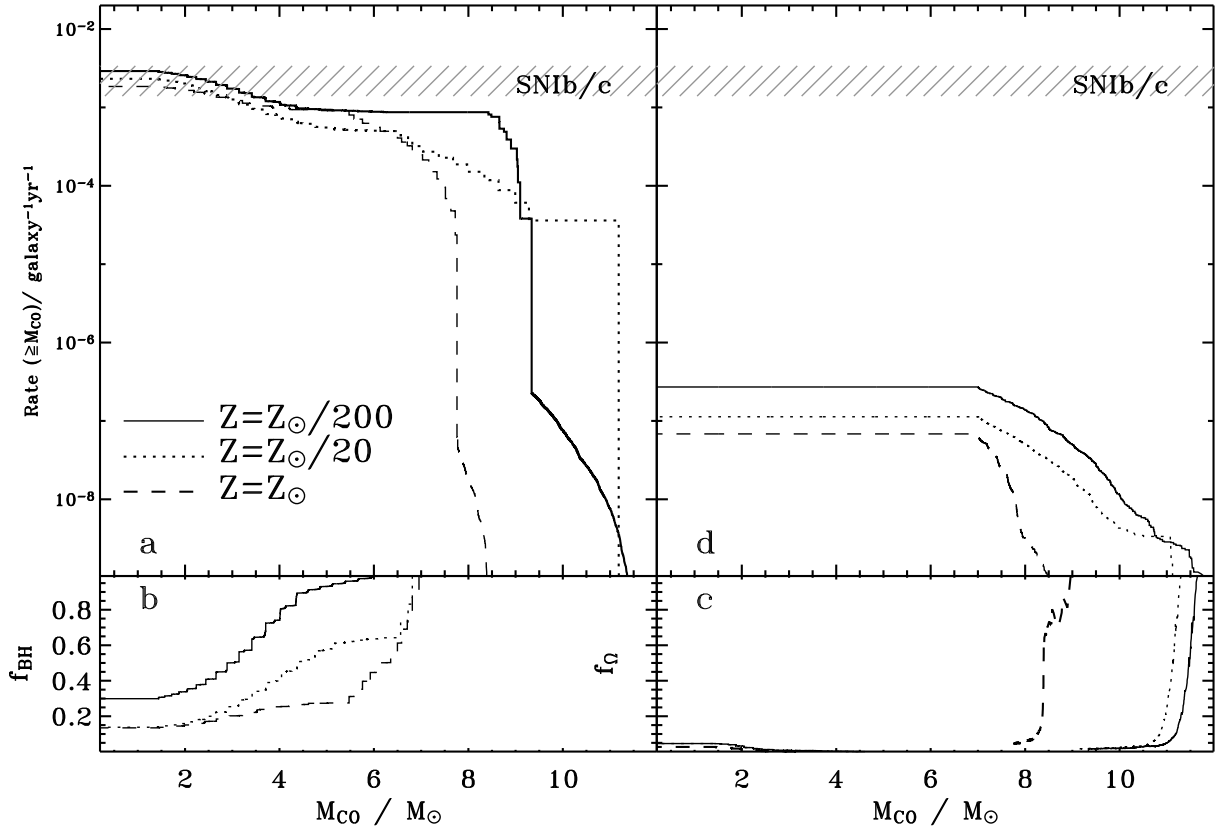,width=\textwidth}} {\caption{Similar
to Fig. \ref{fig11} but with an initial KTG-KTG distribution of
binaries.}
\label{fig12}}
\end{figure*}

\begin{figure*}
\centerline{\psfig{figure=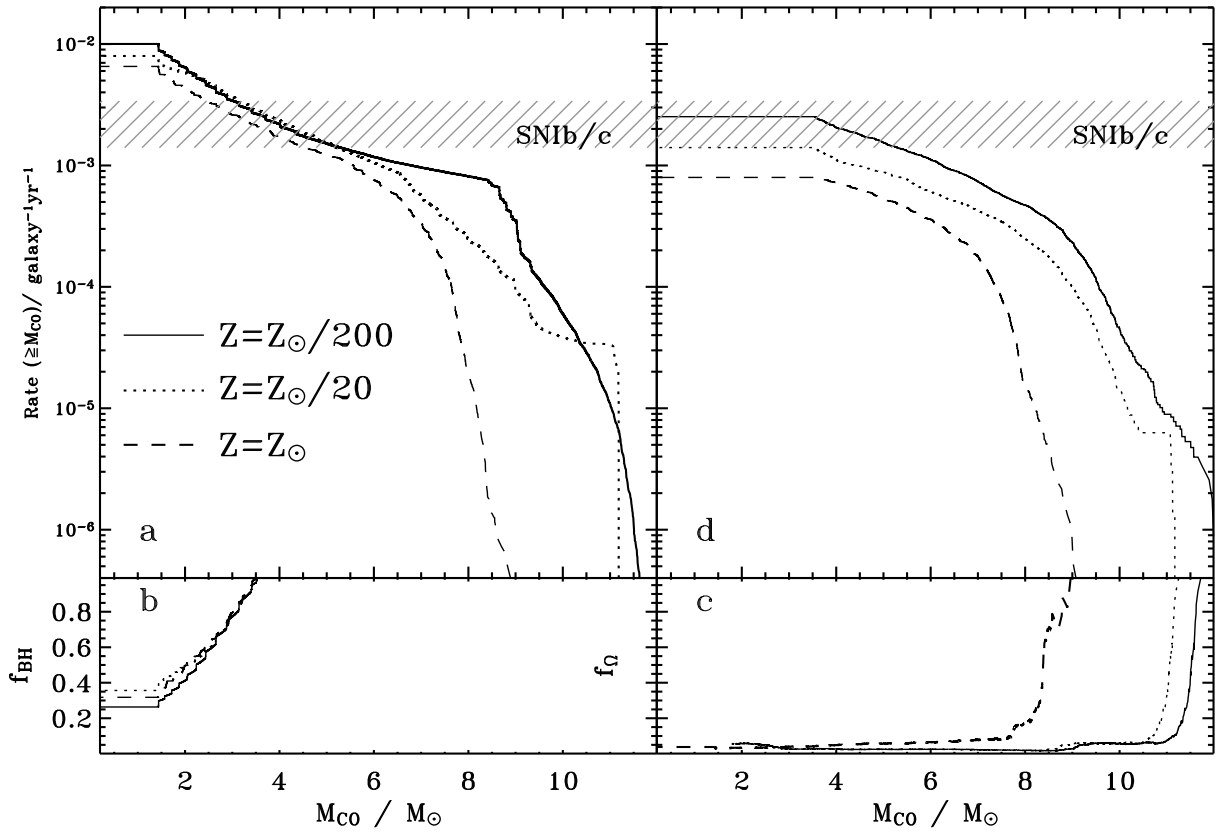,width=\textwidth}} {\caption{Similar
to Fig. \ref{fig11} but using the B02 prescription for NS and BH
masses rather than the H00 formulae.}
\label{fig13}}
\end{figure*}

From our large binary population models with a variety of initial
masses and metallicities we find that both the fraction of stars that
form a central black hole and those that end their lives with
sufficient angular momentum to form a disc significantly increase with
decreasing metallicity (Figs. \ref{fig11} -- \ref{fig13}). This is
because low metallicity keeps the radius of the star smaller and
reduces mass loss. Both properties inhibit the loss of angular
momentum (MacFadyen \& Woosley 1999; Heger \& Woosley 2002) so low-$Z$
stars are likely to be rotating more rapidly. Also, the lower the
metallicity, the higher the stellar mass for WR star formation. The
latter increases the mass of the heaviest CO core and favours black
hole formation which in turn could power a are more luminous burst
because more energy can be extracted from the black hole via e.g. the
Blandford-Znajek mechanism (Blandford \& Znajek 1977; Lee et al. 2000;
Wheeler, Meier \& Wilson 2002).

The likely metallicity dependence of both black-hole formation and
rotation suggests that the production of GRBs is likely to affected by
the physical conditions in the local ISM -- at high $Z$ gas opacities
are larger in the outer stellar layers and so more momentum is
transferred by radiation pressure, mass-loss is more intense and 
GRB formation is disfavoured.

What are the effects of a dependence of GRB luminosity on
the metallicity of their progenitors? The most significant is a
potential offset between the true star-formation rate and that traced
by GRBs (Ramirez-Ruiz, Lazzati \& Blain 2002). If GRBs in
low-metallicity environments and in low-mass galaxies are more
luminous then they are likely to be overrepresented in GRB
samples. Low-mass galaxies are likely to have statistically lower
metallicities and thus contain more luminous GRBs than high-mass
galaxies.  Because galaxy mass is expected to build up monotonically
through mergers,  it is possible that the highest-$z$ GRBs could
be systematically more luminous due to the lower mass of their
hosts. This effect is likely to be more significant than, but in the
same direction as, the global increase in metallicity with cosmic
time.

The most luminous GRBs of all could be associated with metal-free
Population-III stars but their very high redshifts would make
examples difficult to find even in the {\it Swift}\footnote{\tt
http://swift.gsfc.nasa.gov} catalogue of hundreds of bursts.

Star-formation activity is likely to be enhanced in merging
galaxies. In major mergers of gas-rich spiral galaxies this
enhancement takes place primarily in the inner kpc, as bar
instabilities drive gas into the core (Mihos \& Hernquist
1994). Metallicity gradients in the gas are likely to be smoothed out
both by mixing prior to star formation and by SN enrichment during
the burst of activity. GRB luminosities could be suppressed in
such well-mixed galaxies making GRBs more difficult to detect in
these most luminous objects. 

Shocks in tidal tails associated with merging galaxies are also likely
to precipitate the formation of high-mass stars, yet as such tails are
likely to consist of relatively low-metallicity gas, it is perhaps
these less intense sites of star-formation at large distances from
galactic radii that are more likely to yield detectable GRBs. This
might have the unfortunate consequence of making GRBs more difficult
to use as clean markers of high-$z$ star-formation activity. More
optimistically, the astrophysics of star formation in high-redshift
galaxies could be studied using the intrinsic properties of a
well-selected population of GRBs with deep, resolved host galaxy
images. If there is a bias towards the discovery of GRBs in
low-metallicity regions, then the GRB host galaxy luminosity function
will be biased to low luminosities.

\section{Comparison with observations}

The high Lorentz factors and energies seen in GRBs are consistent with
the catastrophic formation of a stellar black hole with mass of a few
$M_\odot$, with about  1 \% of the rest mass energy going to a
relativistic outflow.  This could be the extreme example of the
asymmetric explosion produced by supernovae (Khokhlov et al. 1999) in
which, instead of halting at the neutron star stage, the collapse
continues to the black-hole stage producing an even faster jet in the
process (MacFadyen \& Woosley 1999). GRBs arising from a small
fraction of stars that undergo this type of catastrophic energy
release are likely to produce collimated outflows.

Even if the outflow is not highly collimated some beaming is expected
because energy would channeled preferentially along the rotation
axis. Also, one would expect baryon contamination to be lowest near
the axis because angular momentum flings material away from the axis
and material with low-angular momentum falls into the black hole. The
dynamics, however, are complex. While numerical simulations of
collapse scenarios can address the fate of the bulk of the matter
(MacFadyen \& Woosley 1999; MacFadyen, Woosley \& Heger 2001; Aloy et
al. 2000, 2002; Zhang et al. 2003), higher resolution simulations of
the outer layers of the stellar mantle seem to be required because
even a very small number of baryons polluting the outflow could
severely limit the attainable Lorentz factor. The entrained baryonic
mass would need to be below $10^{-4} M_\odot$ to allow these high
relativistic expansion speeds.

Although jets in GRBs were first suggested for GRB 970508 (Waxman,
Kulkarni \& Frail 1998), they were widely invoked for GRB 990123 to
explain its spectacular energy release. Subsequent multi wavelength
observations of GRBs have been interpreted as evidence for explosions
with jet-like geometry (Stanek et al. 1999; Harrison et al. 1999;
Castro-Tirado et al. 1999). The detection of polarization (e.g.,
Covino et al. 1999; Wijers et al. 1999) gave further credence to the
jet hypothesis.   

\begin{figure*}
\centerline{\psfig{figure=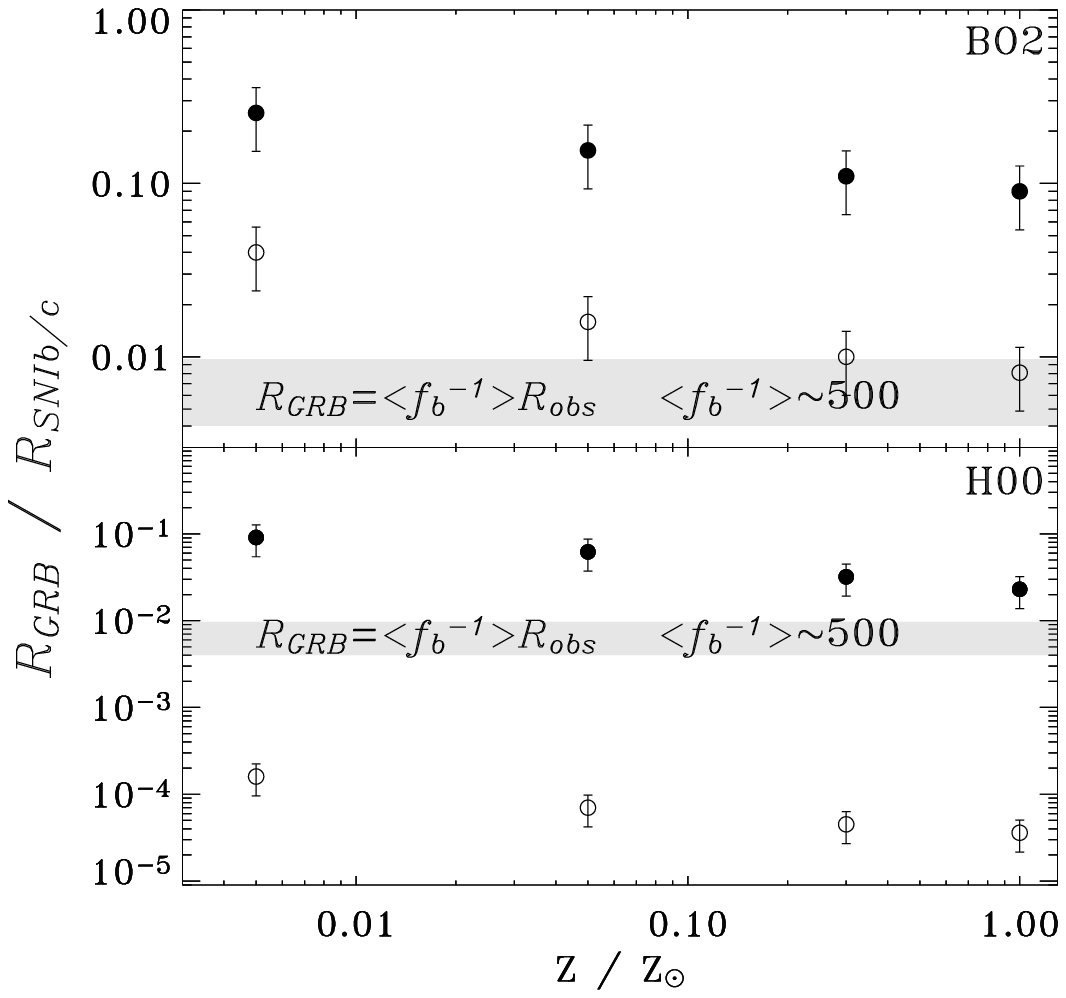,width=\textwidth}} {\caption{GRB
formation rates as a function of metallicity for two different
prescriptions for NS and BH masses. A GRB progenitor is thought to be
a star that has both a massive enough core to form a BH and adequate
rotation rate at the time of collapse to allow the formation of a
centrifugally supported disc (see Section 3). The shaded region marks
the 1$\sigma$ width of the birth rate of GRBs as derived by Frail et
al. (2001). The filled circles are KTG-q binary populations, the empty
circles are binaries with an initial KTG-KTG distribution.}
\label{fig14}}
\end{figure*}

Because conical fireballs are visible to only a fraction $f_b$ of
observers, the true GRB rate is $R_t\; =\; <f_b^{-1}> \;R_{\rm obs}$
where $R_{\rm obs}$ is the observed GRB rate and $<f_b^{-1}>$ is the
mean of the beaming fractions. Frail et al. (2001) recently derived
$<f_b^{-1}> \approx 500$ for a comprehensive sample of GRB afterglows
with known distances based on observed broadband breaks in their light
curves\footnote{This estimate assumes that the breaks observed in many
GRB afterglow lightcurves are due to a geometrical beam effect and not
to either a transition to non-relativistic expansion or an
environmental effect such as a sharp density gradient. An important
feature produce by this jet collimation is the achromaticity of the
afterglow break, which clearly distinguishes it from the steepening
that may be produced by the passage of a spectral break through the
observing band.}. This implies, within the uncertainties and possible
limitations of such a method, that only a small fraction of GRBs are
visible to a given observer and therefore the true GRB rate is several
hundred times larger than the observed one: $R_{\rm obs} \approx
10^{-5} R_{\rm SNIb/c}$ (Schmidt 2001). Fig. \ref{fig14} shows our
formation rates (as a function of $Z$) for GRB progenitors, stars that
have both a massive enough core to form a black hole and adequate
rotation at the time of collapse to allow the formation of a
disc. Clearly these rates are highly sensitive to the assumed black
hole formation model. Binary models in the KTG-q case are easily
capable of supplying a sufficient number of progenitors even if the
GRB rate is several hundred times larger than the observed rate. A
discrepancy lies in the birth rates of collapsars when using an
initial KTG-KTG distribution of binaries because the theoretical rates
in the H00 case are low but not for the B02 case which are in good
agreement. This can be reconciled by assuming a smaller mass threshold
for black hole formation (see equation \ref{mco}).

\section{Conclusions}

The primary purpose of this paper was to use the rapid
binary-evolution algorithm to evaluate the formation rate of
interesting individual species of WR stars that may be the progenitors
of the common long-soft GRBs. Rotating naked helium stars, presumed to
have lost their envelopes in winds or to companions, are followed from
the ZAMS up to the formation of the CO core. Recognizing that the two
essential ingredients for the collapsar model are a sufficiently
massive core to form a black hole and enough rotation to form a disc,
we have studied their effects on the results of binary populations
synthesis and made a global comparison with GRB observations.

The framework we have used to determine a star's angular momentum is a
simple one. While many of the processes involved are not understood in
detail, we do have a qualitative picture of how binaries evolve and we
hope to construct a model that correctly follows them through the
various phases of evolution. This method has led us to conclude that
the resulting spin rates for single stars become too low to form a
centrifugally supported disc to drive a GRB engine. Heger \& Woosley
(2002), including more realistic estimates of angular momentum
transport, argue that some single stars may still be able to form a
disc. However inclusion of magnetic torques in their calculations also
results in too little angular momentum for collapsars. Either the
stellar model description is inaccurate or some other evolutionary
path must be involved in making GRBs. The obvious option is a close
binary or merger (see also Fryer et al. 1999), where tidal interaction
transfers orbital angular momentum to the stellar rotation. An
important ingredient is the fact that the rapidly rotating new black
holes that power the GRBs are typically accompanied by MS stars and so
are likely donors of soft X-ray binaries (only those MS stars which
remained gravitationally bound after the explosion).

For most of the individual binary populations we find that the
expected formation rates of collapsars in the Galaxy are easily
capable of supplying a sufficient number of progenitors, even if the
true GRB rate is several hundred times larger than the observed rate
because of beaming. An interesting conclusion of this work is that
binary stars at low metallicity are important for the formation of a
rapidly rotating, massive helium core at collapse. This effect could
increase the GRB formation rates by a factor of 5--7 at
$Z=Z_\odot/200$. We finally note that the simple binary-evolution
algorithm provides an adequate description of the observations
although more stringent constraints to many of the evolution variables
are needed in order to draw more accurate conclusions.

\section*{Acknowledgments}

We are grateful to M.~J. Rees, P.  Podsiadlowski, A. MacFadyen and
J. Hurley for helpful conversations.  We also thank the referee,
K. Belczynski, for comments that led to improvements of this
paper. ER-R acknowledges patronage from CONACYT, SEP and the ORS
foundation. RGI thanks PPARC for support. CAT thanks Churchill College
for a fellowship.

\newpage

\begin{table*}
\vspace{3cm}
{\centering \begin{tabular}{|c|c|c|}
\hline 
\hline 
\( Z=0.02=Z_{\odot } \)&
\textbf{SNIb/c}&
\textbf{SNII}\\
\hline 
\hline 
\textbf{Single Stars}&
\( 2 \times 10^{-3}\pm 1 \)&
\( 17\times 10^{-3}\pm 4 \)\\
\hline 
\multicolumn{1}{|c|}{\textbf{KTG - KTG}}&
\( 1 \times 10^{-3}\pm 1 \)&
\( 11 \times 10^{-3} \pm 4  \)\\
\textbf{KTG - q}&
\( 6 \times 10^{-3}\pm 2 \)&
\( 10 \times 10^{-3} \pm 4 \)\\
\hline 
\end{tabular}\par}
\vspace{0.3cm}
\caption{\label{rates_Z0.02}Model rates of type Ib/c and type II SNe
(per Galaxy per year) for single stars and binaries with the secondary
chosen either from the KTG or q IMF. SN rates include binaries which
may have coalesced or broken up. The associated uncertainties in the
observed offsets represent the 1$\sigma$ confidence region as
calculated from the Poisson errors.  \protect\( S=7.608\,
\textrm{Galaxy}^{-1}\, \textrm{yr}^{-1}\protect \) is the rate of star
formation.}
\end{table*}

\begin{table*}
\vspace{3cm}
{\centering \begin{tabular}{|c|c|c|}
\hline 
\textbf{Stellar Type}&
\textbf{KTG / per cent}&
\textbf{q / per cent}\\
\hline 
\hline 
Low Mass Main Sequence \( M<0.7\, \textrm{M}_{\odot } \)&
\( 77\pm 0.5 \)&
\( (9.0 \pm 0.1)\times 10^{-1} \)\\
\hline 
Main Sequence \( M>0.7\, \textrm{M}_{\odot } \)&
\( 24 \pm 0.1 \)&
\( 95 \pm 0.4 \)\\
\hline 
Hertzsprung Gap&
\( (1.2 \pm 0.2)\times 10^{-3} \)&
\( (1.3 \pm 0.2)\times 10^{-1} \)\\
\hline 
Core Helium burning&
\( (4.4 \pm 0.3)\times 10^{-4} \)&
\( (4.3 \pm 0.3)\times 10^{-1} \)\\
\hline 
Naked Helium Star MS&
\( (2.5 \pm 0.1)\times 10^{-3} \)&
\( 1.2 \pm 0.1 \)\\
\hline 
Naked Helium Star HG&
\( (7.6 \pm 0.3)\times 10^{-3} \)&
\( 1.4 \pm 0.1 \)\\
\hline 
Carbon-Oxygen White Dwarf&
\( (2.8 \pm 0.9)\times 10^{-3} \)&
\( (1.2 \pm 0.5)\times 10^{-1} \)\\
\hline 
Oxygen-Neon White Dwarf&
\( (3.9 \pm 1.2)\times 10^{-3} \)&
\( (9.9 \pm 3.4)\times 10^{-2} \)\\
\hline 
Neutron Star&
\( (9.5 \pm 0.4)\times 10^{-3} \)&
\( (7.5 \pm 0.4)\times 10^{-1} \)\\
\hline 
Black Hole&
\( (2.6 \pm 0.2)\times 10^{-4} \)&
\( (1.3 \pm 0.1)\times 10^{-1} \)\\
\hline 
NS/BH combined&
\( (9.7 \pm 0.5)\times 10^{-3} \)&
\( (8.8 \pm 0.6)\times 10^{-1} \)\\
\hline 
\end{tabular}\par}\vspace{0.3cm}

\caption{\label{companion_stellar_types}Stellar type of the companion
star when the primary explodes as a possible GRB (percentages) for
\protect\( Z=Z_{\odot }\protect \) in the binary phase.}
\end{table*}

\newpage

\newpage

\newpage

\newpage

\newpage

\newpage

\newpage

\newpage

\newpage

\newpage

\newpage

\newpage

\newpage

\newpage

\begin{thebibliography}{}

\bibitem{}Aloy M.~A., Ibanez J.~M., Marti J.~M., Muller E., MacFadyen
A.~I., 2000, ApJ, 531, L119

\bibitem{}Aloy M.~A., Ibanez J.~M., Miralles J.~A., Urpin V., 2002,
A\&A, 396, 693

\bibitem{} Belczynski K., Bulik T., Rudak B., 2002a, ApJ, 571, 394

\bibitem{} Belczynski K., Kalogera V., Bulik T., 2002b, ApJ, 572, 407
(B02)

\bibitem{} Blandford R. D., Znajek R., 1977, MNRAS, 179, 433

\bibitem{} Bombaci I., 1996, A\&A, 305, 871 

\bibitem{} Cameron A.G.W., Mock M., 1967, Nature, 215, 464

\bibitem{} Cappellaro E., Evans R., Turatto M., 1999, A\&A, 351, 459

\bibitem{} Cappellaro E., 2001, Mem. S. A. It., vol 72-4, 863

\bibitem{} Castro-Tirado A.~J. et al., 1999, Science, 283, 5410 

\bibitem{} Costa E. et al., 1997, Nature, 387, 783

\bibitem{} Covino S. et al., 1999, A\&A, 348, L1

\bibitem{} de Jager, C., Nieuwenhuijzen, H. \& van der Hucht, K.,
1988, A\&AS, 72, 259

\bibitem{} Eggleton P.P., Fitchett M., Tout C.A., 1989, ApJ, 347, 998

\bibitem{} Frail D. A., Kulkarni S. R., Nicastro S. R., Feroci M.,
Taylor G. B., 1997, Nature, 389, 271

\bibitem{} Frail D. A. et al., 2001, 562, L55

\bibitem{} Froning C. S., Robinson E. L., 2001, AJ, 121, 2212 

\bibitem{} Fryer C.~L., Woosley S.~E., 1998, ApJ, 502, L9

\bibitem{} Fryer C.~L., Woosley S.~E., Hartmann D.~H., 1999, ApJ, 526, 152
\bibitem{} Fryer C. L., Kalogera V., 2001, ApJ, 554, 548

\bibitem{} Fryer C.~L., Woosley S.~E., Heger A., 2001, ApJ, 550, 372

\bibitem{} Goldberg D., Mazeh T., 1994, A\&A, 282, 801

\bibitem{} Hamann W.-R., Koesterke L., 1998, A\&A, 335, 1003

\bibitem{} Han Z., 1998, MNRAS, 296, 1019

\bibitem{} Harrison F. A. et al., 1999, ApJ, 523, L121

\bibitem{} Heger A., Woosley S.~E., 2002, astro-ph/0206005

\bibitem{} Hurley J.~R., 2000, Ph.D. thesis, Cambridge Univ

\bibitem{} Hurley J.~R.,  Pols O.R., Tout C.A., 2000, MNRAS, 315, 543

\bibitem{} Hurley J.~R., Tout C.A., Pols O.R., 2002, MNRAS, 329, 897
 
\bibitem{} Iben I.Jr, Tutukov A.~V., 1984, ApJS, 54, 335

\bibitem{} Iben I.Jr., Livio M., 1993, PASP, 105, 1373

\bibitem{} Khokhlov A. M. et al., 1999, ApJ, 529, L107

\bibitem{} Klu\'{z}niak W., Lee W. H., 1998, ApJ, 494, L53

\bibitem{} Kroupa P., Tout C.A., Gilmore G., 1993, MNRAS, 262, 545

\bibitem{} Kudritzki R.P., Reimers D., 1978, A\&A, 70, 227 

\bibitem{} Kudritzki R.P., Pauldrach A., Puls J., Abbott D.C., 1989, A\&A, 
219, 205 

\bibitem{} Lang K.R., 1992, Astrophysical Data, Springer-Verlag

\bibitem{} Larson R.~B., 2001, in Zinnecker, H. and Mathieu
R. D. eds., IAU 200 Symposium

\bibitem{} Lee H.~K., Wijers R.~A.~M.~J., Brown G.~E., 2000, PhR, 325,
83

\bibitem{} Lee H.~K., Brown G.~E., Wijers R.~A.~M.~J., 2002, ApJ, 575, 996

\bibitem{} Lee W. H., Ramirez-Ruiz E., 2002, ApJ, 577, 893

\bibitem{} Livio M., Pringle J.~E., 1998, ApJ, 505, 339

\bibitem{} Maeder A., 1991, A\&A, 242, 93

\bibitem{} MacFadyen A.\,I., Woosley S.\,E., 1999, ApJ, 524, 262

\bibitem{} MacFadyen A.~I., Woosley S.~E., Heger A., 2001, ApJ, 550, 410

\bibitem{} Matzner C. D., 2003, MNRAS, 345, 575

\bibitem{} Mazeh T., Goldberg D., Duquennoy A., Mayor M., 1992, ApJ, 401, 265

\bibitem{} McClintock J. E., Garcia M. R., Caldwell N., Falco E. E.,
Garnavich P. M., Zhao P., 2001, ApJ, 551, 147

\bibitem{} M\'esz\'aros P., 2001, Science, 291, 79

\bibitem{} \Mesz P., Rees M.~J., 2001, ApJ, 2001, ApJ, 556, L37 

\bibitem{} \Mesz P.,Waxman E., 2001, Phys. Rev. Lett., 87, 1102

\bibitem{} Mihos J.~C., Hernquist L., 1994, 431, L9

\bibitem{} Nieuwenhuizen H., de Jager C., 1990, A\&A, 231, 134 

\bibitem{} Orosz J. A. et al., 2001, ApJ, 555, 489 

\bibitem{} Phillips J. P., in Torres-Peimbert S. ed., Planetary
Nebulae. IAU 131 Symposium, (Kluwer:Dordrecht), 425

\bibitem{} Pols O.R., Marinus M., 1994, A\&A, 288, 475

\bibitem{} Portegies Zwart S.F., Verbunt F., 1996, A\&A, 309, 179

\bibitem{} Ramirez-Ruiz E., Dray L., Madau P., Tout C.~A., 2001,
MNRAS, 327, 829
 
\bibitem{} Ramirez-Ruiz E., Lazzati D., Blain A.~W., 2002, ApJ, 565, L9

\bibitem{} Ramirez-Ruiz E., Celotti A., Rees M.~J., 2002, MNRAS, 337,
1349

\bibitem{} Rappaport S., Verbunt F., Joss P., 1983, ApJ, 275, 713

\bibitem{} Reg\H{o}s E., Tout C.A., 1995, MNRAS, 273, 146

\bibitem{} Ruffert M., Janka H.-T., 1999, A\& A, 344, 573

\bibitem{} Schmidt M., 2001, ApJ, 552, 36

\bibitem{} Soderberg A. M., Ramirez-Ruiz E., 2002, MNRAS, 330 L24

\bibitem{} Spruit H. C., Phinney E. S., 1998, Nature, 393, 193

\bibitem{} Spruit H. C., 2002, A\&A, 381, 923

\bibitem{} Stanek K. Z., Garnavich P. M., Kaluzny J., Pych W.,
Thompson I.,  1999, ApJ, 522, L39

\bibitem{} Thompson C., 1994, MNRAS, 270, 480

\bibitem{} Tout C.A., Aarseth S.J., Pols O.R., Eggleton P.P., 1997, MNRAS,
           291, 732

\bibitem{} Usov V.~V., 1992, Nature, 357, 472

\bibitem{} van Paradijs J., 1995, in Lewin W.H.G., van Paradijs J.,
van den Heuvel E.P.J., eds., X-ray Binaries.  Cambridge Astrophysics
Series, 26, p.~520

\bibitem{} van Paradijs et al., 1997 Nature, 386, 686

\bibitem{} Vassiliadis E., Wood P.R., 1993, ApJ, 413, 641

\bibitem{} Wagner R. M., Foltz C. B., Shahbaz T., Casares J., Charles
P. A., Starrfield S. G., Hewett P., 2001, ApJ, 556, 42

\bibitem{} Waxman E., Kulkarni S. R., Frail D. A., 1998, ApJ, 497, 288

\bibitem{} Wheeler J. G., Yi I., Hoflich P., Wang L., 2000,  ApJ, 537, 810

\bibitem{} Wheeler J.~C., Meier D.~L., Wilson J.~R., 2002, ApJ, 568, 807

\bibitem{} Whyte C.A., Eggleton P.P., 1985, MNRAS, 214, 357

\bibitem{} Wijers R.~A.~M.~J. et al., 1999, ApJ, 523, L33 

\bibitem{} Woosley S. E., 1986, in Nucleosynthesis and Chemical
Evolution, ed. B. Hauck et al. (Geneva: Geneva Obs.), 1

\bibitem{} Woosley S.\,E., 1993, ApJ, 405, 273

\bibitem{} Zhang W., Woosley S.\,E., MacFadyen A.\,I., 2003, ApJ, 586, 356

\end{thebibliography}
\end{document}